%% file: data-driven_output_regulation_parabolic_PDE.tex
\documentclass[twocolumn]{autart}    
\usepackage{hyperref}

\usepackage{graphicx}          

\usepackage{color}
\usepackage{tikz}
\usepackage{pgfplots}
\usetikzlibrary{fit,calc}
\pgfplotsset{compat=1.14}
\usepackage{psfrag}
\usepackage{amssymb}
\usepackage{graphicx}
\usepackage{amsmath,amsfonts,amsxtra}
\usepackage{mathrsfs}
\usepackage{todonotes}
\usepackage{mathtools,leftidx}
\usepackage[inline]{enumitem}
\usepackage{float}
\usepackage{cases} 
\usepackage{environ}
\usepackage{bm} 
\usepackage{bbm}

\allowdisplaybreaks[1]

\newcommand{\col}[1]{\operatorname{col}(#1)}

\newcommand{\diag}[1]{\operatorname{diag}(#1)}

\renewcommand{\d}{\mathrm{d}}
\renewcommand{\t}{^{\top}}
\newcommand{\tint}{\textstyle\int}

\newcommand{\rk}{\operatorname{rank}}

\makeatletter
\newsavebox{\measure@tikzpicture}
\NewEnviron{scaletikzpicturetowidth}[1]{%
  \def\tikz@width{#1}%
  \begin{lrbox}{\measure@tikzpicture}%
  \BODY
  \end{lrbox}%
  \pgfmathparse{#1/\wd\measure@tikzpicture}%
  \BODY
}

\newcommand\getwidthofnode[2]{%
	\pgfextractx{#1}{\pgfpointanchor{#2}{east}}%
	\pgfextractx{\pgf@xa}{\pgfpointanchor{#2}{west}}
	\addtolength{#1}{-\pgf@xa}%
}
\newcommand\getheightofnode[2]{%
	\pgfextracty{#1}{\pgfpointanchor{#2}{north}}%
	\pgfextracty{\pgf@xa}{\pgfpointanchor{#2}{south}}
	\addtolength{#1}{-\pgf@xa}%
}
\makeatother

\usepackage{tabularx}
\usepackage{todonotes}
\usepackage{boldline}

\newlength\figureheight
\newlength\figurewidth

\definecolor{matlabOrange}{rgb}{0.85000,0.32500,0.09800}%
\definecolor{matlabBlue}{rgb}{0.00000,0.44700,0.74100}%
\definecolor{matlabGreen}{rgb}{0.4660 0.6740 0.1880}

\DeclareRobustCommand\matlabBlueCirc{\tikz{\draw[color=matlabBlue,very thick] (0,0) circle [radius=2.5pt];}}
\DeclareRobustCommand\matlabGreenCirc{\tikz{\draw[color=matlabGreen,very thick] (0,0) circle [radius=2.5pt];}}
\DeclareRobustCommand\matlabOrangeCross{\tikz{\draw[color=matlabOrange,very thick] (0,0) -- (0.15,0.15);\draw[color=matlabOrange,very thick]
 (0,0.15) -- (0.15,0);}}

\DeclareRobustCommand\matlabBlueLine{\tikz[baseline=-0.6ex]{\draw[color=matlabBlue,very thick] (0,0) -- (0.25,0);}} 
\DeclareRobustCommand\matlabOrangeLine{\tikz[baseline=-0.6ex]{\draw[color=matlabOrange,very thick] (0,0) -- (0.1,0);\draw[color=matlabOrange,very thick]
 (0.15,0) -- (0.25,0);}}
 
\begin{document}

\begin{frontmatter}

\title{A Koopman-backstepping approach to data-driven robust output regulation for linear parabolic systems} 


\author{Joachim Deutscher}\ead{joachim.deutscher@uni-ulm.de},
\author{Julian Zimmer}\ead{julian.zimmer@uni-ulm.de}  

\address{Institute of Measurement, Control and Microtechnology, Ulm University, 
Germany}


\begin{abstract}
In this paper a solution of the data-driven robust output regulation problem for linear parabolic systems is presented. Both the system as well as the ODE, i.e., the disturbance model,  describing the disturbances are unknown, but finite-time sequential data obtained from measurements of the output to be controlled and additional boundary outputs are available. The data-driven controller is designed in the Koopman operator framework for PDEs, where the Koopman modes and eigenvalues are obtained from data using Hankel-DMD. It is shown that all system parameters and the eigenvalues of the disturbance model can be recovered from the available measurements by solving an inverse Sturm-Liouville problem. This allows to directly apply backstepping methods for the robust regulator design. For this, closed-loop stability in the presence of small errors in the Hankel-DMD is verified in the nominal case. Robust output regulation is shown for non-destabilizing model uncertainties. A numerical example demonstrates the results of the paper. 
\end{abstract}
\end{frontmatter}

\section{Introduction}
The use of data for the analysis and control of dynamical systems receives currently a growing interest in the control community (see \cite{Br22}). This is attributed to the large availability of data from measurements or simulation of complex systems. The main motivation for data-driven control is to circumvent a possibly time consuming system identification (see \cite{Lj99}) or to control complex systems (see \cite{Br22}), where a first-principle modelling is very difficult. Since the new methods are not based on a system description in terms of differential equations, they also offer new research possibilities for the system identification of complex systems including models described by nonlinear ODEs and PDEs. Of particular interest in this direction is the application of the \emph{Koopman operator} (see \cite{Ma20}). This approach describes the nonlinear system dynamics in the space of observables (i.e., generalized measurements), where their time evolution is described by the linear Koopman operator. Hence, nonlinear systems can be identified by solving a linear regression problem for the Koopman operator (see \cite{Mau20a}). A similar situation also arises for distributed-parameter systems, where the system identification is mainly based on finite-dimensional approximations (see \cite{Pol76,Kub77,Ba98} for an overview). Only a few results exist, which directly use the distributed-parameter system for identification without an approximation (see, e.g., \cite{Ru08a,Kn13,Gehr16}). Recently, the Koopman framework was extended to PDEs in \cite{Na20}, where a systematic spectral analysis of the Koopman operator for PDEs is presented. These new results were already applied in \cite{Mau21} to solve identification problems for nonlinear PDEs using state data and for the data-driven control of linear parabolic PDEs in \cite{Deu24}. In \cite{Deu24a} the Koopman operator was applied to solve a data-driven robust output regulation problem for finite-dimensional LTI systems. This amounts to design a regulator ensuring the tracking of reference inputs in the presence of disturbances and model uncertainties. Both the unknown system and the unknown disturbance model are determined from measurement data. For an unknown signal model and a known parabolic PDE, output regulation has been achieved using adaptive control under the condition of persistency of excitation (see \cite{Gu022} and the references therein). To the best know\-ledge of the authors, however, a fully data-driven solution of the robust output regulation problem for distributed-parameter systems with unknown plant and disturbance model has not been obtained in the literature so far. This suggests to use the results for finite-dimensional systems in \cite{Deu24} and the new methods in \cite{Na20} for the Koopman operator related to PDEs in order to solve a data-driven robust output regulation problem for parabolic systems. 

In this paper a data-driven solution of the robust output regulation problem for linear parabolic systems with boundary control is presented by combining the Koopman theory with the backstepping approach (see \cite{Kr08}). Both the system and the disturbance model are unknown, but only a finite-time sequence of measurements for the output to be controlled and of additional boundary measurements is available. By extended the Hankel-DMD to parabolic PDEs (see \cite{Ar17,Dr18,Me22} for ODEs), the Koopman modes and eigenvalues of the Koopman operator related to the PDE and the disturbance model are determined from data. Usually, this information is used to obtain a finite-dimensional surrogate model in Koopman modal coordinates for data-driven control (see, e.g., \cite{Kai21,Gos21,Deu24a} for ODEs). This, however, prevents the application of late-lumping methods (see, e.g., \cite{Aul16}) and in particular backstepping for the regulator design. In order to be able to use the latter approach, it is shown that two Koopman modes and eigenvalues of the parabolic system are sufficient to solve an \emph{inverse Sturm-Liouville problem} (see \cite{Kra20}). Hence, choosing two dominant modes of the parabolic PDE allows to accurately identify all parameters, since these modes are well represented in the utilized sequential data. Another advantage of this method compared to other late lumping identification methods in \cite{Ru08a,Kn13,Gehr16} lies in the fact that the vectors describing the disturbance input locations have not to be determined. This results in a systematic method for the identification of the system parameters and the eigenvalues of the disturbance model. The latter are sufficient to define an internal model for the design of a robust regulator (see \cite{Deu16a,Deu20}).  This result is also of interest for the recently introduced framework of \emph{neural operators} for backstepping (see \cite{Kr24}), which allows to design a backstepping controller without solving the kernel equations. In particular, the identified parameters can directly be applied to obtain the backstepping gains from the neural operator to quickly accommodate for varying parameters and disturbance signal forms in various identification phases. 

The next section introduces the considered data-driven robust output regulation problem. Then, the Koopman operator model for the parabolic system and the disturbances is derived in Section \ref{sec:koop}. For determining the Koopman modes and eigenvalues from data, the Hankel-DMD is extended to distributed-parameter systems in Section \ref{sec:ddsolkoop}. In addition, also a numerical reliable implementation in form of an SVD-enhanced DMD is provided. By solving an inverse Sturm-Liouville problem it is shown that only two Koopman modes and eigenvalues of the parabolic PDE are sufficient to identify all its parameters with small errors. In Section \ref{sec:sfr} the regulator design presented in \cite{Deu16a,Deu20} is applied to the identified parabolic system and the disturbance model in the nominal case. Closed-loop stability is verified for small errors in the Hankel-DMD. With this, robust output regulation is shown in Section \ref{sec:robreg}. A numerical example demonstrates the results of the paper, which also provides a systematic for choosing the sampling time and the number of samples to determine the measurement data.

\textit{Notation.} If convenient the $(z,t)$-dependence is dropped for clarity, i.e., $x(0,t) = x(0)$, $x(z,t) = x$, for example. The notations $\dot{x}(z,t) = \partial_tx(z,t)$ and $x''(z,t) = \partial^2_zx(z,t)$ are used in the paper.

\section{Problem formulation}\label{sec:probform}
Consider the system described by the linear parabolic PDE
\begin{subequations}\label{plant}
\begin{align}
 \dot{x}(z,t)    &= (\rho + \Delta\rho(z)) x''(z,t) + (a + \Delta a(z))x(z,t)\nonumber\\
                &\quad   + g_1\t(z)d(t)\label{parasys}\\
  	  x'(0,t) &= (q_0 + \Delta q_0)x(0,t) + g_2\t d(t),                     &&\hspace{-1.3cm} t > 0\\ 
	  x'(1,t) &= (q_1 + \Delta q_1)x(1,t) + u(t) + g_3\t d(t)               &&\hspace{-1.3cm} t > 0 \label{parasys2}\\
         y(t) &= x(z_0,t) + g_4\t d(t),                                            &&\hspace{-1.3cm} t \geq 0\label{hypout1}\\
      \eta(t) &= \begin{bsmallmatrix} x(0,t) \\ x(1,t) \end{bsmallmatrix} + G_5 d(t),                                       &&\hspace{-1.3cm} t \geq 0,\label{hypout2}
\end{align}%
\end{subequations}
with the state $x(z,t) \in \mathbb{R}$ defined on $(z,t) \in [0,1]\times\mathbb{R}^+$, the input $u(t) \in \mathbb{R}$, the unknown disturbance $d(t) \in \mathbb{R}^{q}$ and the measurement $\eta(t) \in \mathbb{R}^2$. It is assumed that $\rho > 0$ and that $g_1$ is piecewise continuous. The output to be controlled $y(t) \in \mathbb{R}$ is available for measurement and $z_0 \in (0,1)$. This class of distributed-parameter systems describe, for example, the temperature profiles in chemical reactors and battery systems or thermal processes in the steal industry. \emph{All} system parameters $\rho, a, q_i, i= 0,1$, being elements of $\mathbb{R}$, $g_i \in \mathbb{R}^q$, $i = 1,\ldots,4$,  $G_5 \in \mathbb{R}^{2\times q}$ and the \emph{model uncertainties} $\Delta\rho$, $\Delta a$, $\Delta q_0$ and $\Delta q_1$ are \emph{unknown}. It is assumed that $\rho + \Delta\rho(z) > 0$, $z \in [0,1]$, with $\Delta\rho, \Delta a \in C[0,1]$. Only $z_0$ defining the output to be controlled is assumed to be known.  The initial condition (IC) of \eqref{plant} is $x(z,0) = x_0(z) \in \mathbb{R}$. Note that the results for the paper can also be directly applied to Dirichlet boundary conditions at the boundaries. If all model uncertainties are set to zero, then the \emph{nominal system}
\begin{subequations}\label{plantnom}
	\begin{align}
	\dot{x}(z,t) &= \rho x''(z,t) + ax(z,t)\label{parasysn}\\
	x'(0,t) &= q_0x(0,t)        \\ 
	x'(1,t) &= q_1x(1,t) + u(t)\\
	   y(t) &= x(z_0,t)             
	\end{align}
\end{subequations}
results, which is unknown and determined by the proposed data-based approach. With this, the model uncertaintes $\Delta\rho(z)$, $\Delta a(z)$, $\Delta q_0$ and $\Delta q_1$ take into account the errors in the identification and parameter changes during operation.
The disturbance $d(t) \in \mathbb{R}^q$ is generated by the \emph{unknown disturbance model}
\begin{subequations}\label{dmod}
\begin{align}
\dot{\omega}_d(t) &= S_d\omega_d(t),    && t > 0,\ \omega_d(0) = \omega_{d,0} \in \mathbb{R}^{n_d}\label{sds}\\
	         d(t) &= P_d\omega_d(t),    && t \geq 0,\label{dout1}
\end{align}	 
\end{subequations}
where $(P_d,S_d)$ is observable and $S_d \in \mathbb{R}^{n_d \times n_d}$ has simple eigenvalues on the imaginary axis. This includes signal models for constant and sinusoidal disturbances appearing frequently in applications. 
It is assumed that by making use of measurements the sequential \emph{output data}
\begin{equation}
	\mathbbm{Y}_{2n} = \begin{bmatrix}y(0) & y(1)    & \ldots & y(2n-1)\\
	                                 \eta(0) & \eta(1) & \ldots & \eta(2n-1)
	                  \end{bmatrix} \in  \mathbb{R}^{3 \times 2n}\label{ydata} 
\end{equation} 
is collected for the some IC $x(0)$ in the presence of the disturbance $d$ and $u \equiv 0$. Note that due to the boundary actuation, the data does not have to take the input $u$ into account. Therein, the notation $y(k) = y(kt_s)$, $k \in \mathbb{N}_0$, with the \emph{sampling time} $t_s > 0$ is employed. The reference input $r(t) \in \mathbb{R}$ is described by the solution of the  \emph{known reference model}
\begin{subequations}\label{rmod}
	\begin{align}
	\dot{\omega}_r(t) &= S_r\omega_r(t),  &&\hspace{0cm} t > 0,\ \omega_r(0) = \omega_{r,0} \in \mathbb{R}^{n_r}\\
	r(t) &= p\t_r\omega_r(t),             &&\hspace{0cm} t \geq 0,
	\end{align}	 
\end{subequations}
in which $(p\t_r,S_r)$ is observable and $S_r \in \mathbb{R}^{n_r \times n_r}$ has only eigenvalues on the imaginary axis. Hence, the reference inputs can be constant, polynomial and trigonometric functions. In the sequel it is assumed that $r$ is available for the regulator to be determined.%

In this paper a continuous-time \emph{dynamic state feedback regulator} is designed ensuring%
\begin{equation}\label{outreg}
 \lim_{t \to \infty}e_y(t) =  0
\end{equation}
with the output tracking error $e_y(t) = y(t) - r(t)$ for all $x(0) \in \mathbb{R}$, controller IC, $\omega_d(0) \in \mathbb{R}^{n_d}$, $\omega_r(0) \in \mathbb{R}^{n_r}$ and any non-destabilizing model uncertainties $\Delta\rho(z)$, $\Delta a(z)$, $\Delta q_0$ and $\Delta q_1$. For this only the output data \eqref{ydata} and the reference model \eqref{rmod} have to be known.

\section{Koopman operator model}\label{sec:koop}
The autonomous nominal system \eqref{plant} with the modelling of the disturbances by \eqref{dmod} can be described by the \emph{extended system}
\begin{equation}\label{ibvp}
 \dot{x}_e(t) = \mathcal{A}_ex_e(t), \quad t > 0, x_e(0) = x_{e,0} \in  D(\mathcal{A}_e)
\end{equation}
with the state $x_e = \col{\omega_d,x}$ in the state space $X_e = \mathbb{C}^{n_d} \oplus L_2(0,1)$ endowed with the \emph{inner product}
$\langle x, y\rangle = \langle x_1,y_1\rangle_{\mathbb{C}^{n_d}} + \langle x_2,y_2\rangle_{L_2} = x_1\t\overline{y}_1 + \tint_0^1x_2(\zeta)\overline{y_2(\zeta)}\d\zeta$.
Then, the system operator reads $\mathcal{A}_eh = \col{S_dh_1,\rho h_2'' + ah_2 + g_1\t P_d h_1} = \col{\mathcal{A}_1h_1,\mathcal{A}_2h}$ for $h = \col{h_1,h_2} \in D(\mathcal{A}_e) \subset X_e$ with $D(\mathcal{A}_e) = \{h \in \mathbb{C}^{n_d} \oplus H^2(0,1)\; |\; h_2'(0) = q_0h_2(0) + g_2\t P_dh_1,\ h_2'(1) = q_1 h_2(1) + g_3\t P_dh_1\}$. The following lemma asserts the well-posedness of \eqref{ibvp}.
\begin{lem}[Extended system]\label{lem:exsys}
 Assume that the spectrum $\sigma(\mathcal{A}_e)$ of $\mathcal{A}_e$ is simple, then $\mathcal{A}_e$ is a \emph{Riesz spectral operator}. Consequently, the initial boundary value problem \eqref{ibvp} is \emph{well-posed}.
\end{lem}	
\begin{pf}
Consider the linear operator $\mathcal{A}_e + \mu I$, $\mu \in \mathbb{R}$. By making use of a direct calculation it is readily verified that $(\mathcal{A}_e + \mu I)x = y$ always has a unique solution $x \in D(\mathcal{A}_e)$ for any $y \in X_e$ and $(\mathcal{A}_e + \mu I)^{-1}$ is bounded if $0 \notin \sigma(\mathcal{A}_e+\mu I)$. There always exists a $\mu \in \mathbb{R}$ satisfying the latter condition so that $\mathcal{A}_e + \mu I$ is closed (see \cite[A.3.49]{Cu20}) and so is $\mathcal{A}_e$. It is straightforward to verify that the eigenvectors $\{\Phi_i,$ $i \in \mathbb{N}\}$ of $\mathcal{A}_e$ are
\begin{subnumcases}{\Phi_i = \label{vmod}}
	\col{v_i,\phi_i},                            & $i = 1,\ldots,n_d$\label{vvec1}\\                                          
	\col{0,\phi_i},                            & $i > n_d$,\label{vvec2}
\end{subnumcases}
in which $v_i \in \mathbb{C}^{n_d}$ are the eigenvectors of $S_d$ and $\phi_i \in L_2(0,1)$, $i = 1,\ldots,n_d$, result from solving
\begin{subequations}\label{phiu}
\begin{align}
&\rho\phi_i''(z) + a\phi_i(z) + g_1\t(z)P_dv_i = \lambda_i\phi_i(z) \label{parasyse}\\
&\phi_i'(0) = q_0\phi_i(0) + g_2\t P_dv_i\\
&\phi_i'(1) = q_1\phi_i(1) + g_3\t P_dv_i.\label{parasys2e}
\end{align} 	 
\end{subequations}
In what follows, it is assumed that the $\phi_i$, $i > n_d$, have been arranged according to a decreasing real part of the corresponding eigenvalues.
Furthermore, with the operator 
\begin{equation}\label{Adef}
 \mathcal{A}h = \rho h'' + ah, \quad h \in D(\mathcal{A})
\end{equation}
and $D(\mathcal{A}) = \{h \in L_2(0,1) \;|\; h'(0) = q_0h(0),\ h'(1) = q_1h(1)\} \subset L_2(0,1)$, the vectors $\{\phi_i,$ $i > n_d\}$ in \eqref{vvec2} solve the eigenvalue problem $\mathcal{A}\phi_i = \lambda_i\phi_i$, $\phi_i \in D(\mathcal{A})$, $i > n_d$. The operator $-\mathcal{A}$ is a Sturm-Liouville operator (see \cite{Del03}) so that its eigenvectors $\{\phi_i, i > n_d\}$ form an orthonormal Riesz basis for $L_2(0,1)$. By making use of this property, a simple calculation shows that $\{\Phi_i,$ $i \in \mathbb{N}\}$ is an $\omega$-linearly independent sequence, which is quadratically close to an orthonormal basis for $X_e$. Hence, by \emph{Bari's theorem} the sequence $\{\Phi_i,$ $i \in \mathbb{N}\}$ is a Riesz basis for $X_e$ (see \cite[Th. 2.7]{Guo19}). This and the fact that $\sigma(\mathcal{A}_e)$ is simple verifies all properties of a Riesz spectral operator according to \cite[Def. 3.2.6]{Cu20}). Consequently, $\mathcal{A}_e$ is the generator of a $C_0$-semigroup implying well-posedness of \eqref{ibvp} (see \cite[Th. 3.2.8]{Cu20}).
\hfill $\Box$
 \end{pf}
This result, in particular, ensures the unique existence of a  $C_0$-semigroup $\mathcal{T}_e(t): X_e \to X_e$, $t \in \mathbb{R}^+$ (see \cite[Th. 3.2.8]{Cu20}), which is generated by $\mathcal{A}_e$, and that the eigenvectors of $\mathcal{A}_e$ span a Riesz basis for $X_e$. The \emph{Koopman operator} $\mathscr{K}(t)$ related to \eqref{ibvp} describes the time evolution of \emph{observables} $g[x_e]$, $x_e \in X_e$, on the state space $X_e$. They are linear functionals, i.e., $g : X_e \to \mathbb{C}$ being elements of the infinite-dimensional space $\mathscr{O}$ of observables. Since any measurement for \eqref{plant} is a functional, the observables $g[x_e]$ can be seen as generalized measurements for \eqref{ibvp}. The \emph{Koopman operator} related to \eqref{ibvp} is defined by the composition
\begin{equation}\label{Kop}
\mathscr{K}(t)g[x_e] = g[\mathcal{T}_e(t)x_e], \quad t \geq 0,
\end{equation} 
for $g \in \mathscr{O},$ $x_e \in D(\mathcal{A}_e)$, which determines the temporal evolution of the observable $g[x_e]$.  For each $t \in \mathbb{R}^+$ the map $\mathscr{K}(t): \mathscr{O} \to \mathscr{O}$ is a linear infinite-dimensional operator (see \cite{Na20}) and uniquely exists by the uniqueness of the $C_0$-semigroup $\mathcal{T}_e(t)$. Note that even though the observables considered in the paper are linear, this requires to introduce the Koopman operator $\mathscr{K}(t)$ as an evolution operator in view of \eqref{Kop}. 
\begin{rem}	
Note that the definition of $\mathscr{O}$ allows to also use the observable $g[x_e] = \mathcal{C}_{z_0}x$, $z_0 \in [0,1]$, with the unbounded \emph{evaluation operator} $\mathcal{C}_{z_0}h = h(z_0), \quad D(\mathcal{C}_{z_0}) = D(\mathcal{A}_e)$. Since $x_e(t) \in D(\mathcal{A}_e)$, $t > 0$, for $x_e(0) \in D(\mathcal{A}_e)$, these evaluations are well-defined (see \cite[Th. 2.1.13]{Cu20}). Further note that $\mathcal{C}_{z_0}\mathcal{T}_e(t)x_e$, $x_e \in D(\mathcal{A}_e)$, is a bounded linear operator for $t \in (0,\infty)$ (see, e.g., the proof of Theorem  \ref{thm:stabnom}). Hence, the output data \eqref{ydata} consists of valid observables in $\mathscr{O}$.
 \hfill $\triangleleft$	 
\end{rem}	
Consider the functionals $g[x_e]$ contained in $X_e' \subset \mathscr{O}$, i.e., linear bounded functionals. By the Riesz representation theorem (see \cite[Th. A.3.55]{Cu20}) they are of the general form $g[x_e] = \langle x_e,\Gamma\rangle = \langle \omega,\gamma_1\rangle_{\mathbb{C}^{n_d}} + \langle x,\gamma_2\rangle_{L_2}$ with $\Gamma = \col{\gamma_1,\gamma_2}$ as well as $\gamma_1 \in \mathbb{C}^{n_v}$ and $\gamma_2 \in L_2(0,1)$. Let $\Psi_i$ be the eigenvector of the adjoint operator $\mathcal{A}_e^*$, i.e., $\mathcal{A}_e^*\Psi_i = \overline{\lambda}_i\Psi_i$, $\Psi_i \in D(\mathcal{A}_e)$, $i \in \mathbb{N}$. Then, the \emph{eigenvalue problem} for the Koopman operator is given by 
\begin{equation}\label{Kopeig}
\mathscr{K}(t)\varphi_i[x_e] =  \e^{\lambda_i t}\varphi_i[x_e], \quad i \in \mathbb{N},
\end{equation}
with the \emph{Koopman eigenfunctionals} $\varphi_i[x_e] = \langle x_e,\Psi_i\rangle$ w.r.t. the \emph{Koopman eigenvalues} $\lambda_i$ (see \cite{Deu24}). Since the sequence $\{\Phi_i,$ $i \in \mathbb{N}\}$ of eigenvectors for $\mathcal{A}_e$ w.r.t. the eigenvalues $\lambda_i$ are a Riesz basis for $X_e$ (cf. Lemma \ref{lem:exsys}) and $\{\Psi_i,$ $i \in \mathbb{N}\}$ is the related biorthonormal sequence, the Koopman eigenfunctionals already fully characterize the vector-valued observable $x_e$ according to $x_e  =  \sum_{i=1}^{\infty}\langle x_e,\Psi_{i}\rangle \Phi_{i} = \sum_{i=1}^{\infty}\Phi_{i}\varphi_{i}[x_e]$. Therein, the \emph{Koopman modes} $\Phi_{i}\in X_e$, $i \in \mathbb{N}$, of $x_e$ coincide with the eigenvectors of $\mathcal{A}_e$ (cf. \eqref{vmod}). Obviously, the Koopman eigenvalues determine the stability, since they coincide with the eigenvalues of $\mathcal{A}_e$. 

\section{Data-driven Koopman modal analysis}\label{sec:ddsolkoop}
\subsection{Hankel-DMD}\label{subsec:hankeldmd}
In what follows, the \emph{Hankel-DMD} is extended to the PDE-ODE system \eqref{plant} and \eqref{dmod} (for the ODE case see, e.g., \cite{Ar17,Dr18,Me22}). This allows to determine the Koopman modes of the outputs $y$ and $\eta$ as well as the corresponding Koopman eigenvalues from data. To this end, introduce the vector-valued observable
\begin{equation}\label{xobs}
\xi  = \begin{bmatrix}
y\\
\eta
\end{bmatrix}
=
\underbrace{\begin{bmatrix}
	g_4\t P_d &\; \mathcal{C}_{z_0}\\
	G_5 P_d &\; \begin{bsmallmatrix}
	\mathcal{C}_0\\
	\mathcal{C}_1
	\end{bsmallmatrix}
	\end{bmatrix}}_{\mathcal{M}}x_e,
\end{equation}
for which the data \eqref{ydata} is available. Then, the \emph{Hankel matrix}
\begin{equation}\label{Hmat}
H_{n}(\xi) = \begin{bsmallmatrix}
\xi(0)      & \xi(1)   & \ldots & \xi(n-1)\\
\xi(1)      & \xi(2)   & \ldots & \xi(n)\\
\vdots      & \vdots   & \vdots & \vdots\\
\xi(n-1)     & \xi(n) & \ldots & \xi(2n-2)
\end{bsmallmatrix} \!\in\! \mathbb{R}^{3n \times n}
\end{equation}
of depth $n$ containing the samples $\xi(i) = \xi[x_e(it_s)]$ of the observable $\xi$ can be formed from \eqref{ydata}.
The next lemma presents a finite-dimensional approximation $F \in \mathbb{R}^{n \times n}$ of the Koopman operator $\mathscr{K}(t_s)$ in form of a companion matrix. For this, note that $\xi(i+1) = \mathscr{K}(t_s)\xi(i) \in \mathbb{R}^3$, $i \geq 0$, holds.
\begin{lem}[Companion matrix]\label{lem:dmd}
Let $f = \col{f_0,\ldots,\allowbreak f_{n-1}} = -H^\dagger_n(\xi) \col{\xi(n-1),\ldots,\xi(2n-1)} \in \mathbb{R}^{n}$ then with the unit vector $e_n \in \mathbb{R}^n$ it holds
\begin{equation}\label{fdmd}
\mathscr{K}(t_s)H_n(\xi) = H_n(\xi)
	\!\underbrace{\begin{bsmallmatrix}
		0      &  0      & \ldots & 0   & -f_0\\
		1      &  0      & \ldots & 0   & -f_1\\
		0      &  1      & \ldots & 0   & -f_2\\
		\vdots &  \vdots & \ddots & 0   & \vdots\\
		0      &  0      & \ldots & 1   & -f_{n-1}
		\end{bsmallmatrix}}_{F}  +\, r_{n}e_{n}\t
\end{equation}
 and $r_{n} \in \mathbb{R}^{3n}$ has minimal norm. Therein, $H_n^\dagger(\xi)$ denotes the \emph{Moore-Penrose pseudo inverse} (see, e.g., \cite[Ch. 12.8]{La85}).
\end{lem}
\begin{pf}	
The result directly follows from considering the minimization problem $\|r_{n}\| = \|H_{n}(\xi)f +  \col{\xi(n),\ldots,\xi(2n-1)}\| \stackrel{!}{=} \text{min}$ with the usual vector $2$-norm in $\mathbb{R}^{3n}$, which is solved by $f$ in Lemma \ref{lem:dmd}. \hfill $\Box$
\end{pf}	
If $r_n = 0$ in \eqref{fdmd} and
\begin{equation}\label{AB}
 x_e(0) = \sum_{i=1}^n\Phi_i\varphi_i[x_e(0)], \quad n > n_d,
\end{equation}
with $\varphi_i[x_e(0)] \neq 0$, $i = 1,\ldots,n$, are fulfilled (see next section for a detailed discussion on exact Hankel-DMD), then the span of the columns in $H_n(\xi)$ is invariant w.r.t. the action of the Koopman operator $\mathscr{K}(t_s)$ so that in this case $F$ is a finite-dimensional representation of the Koopman operator on this subspace. Hence, for $r_n \neq 0$ the matrix $F$ can be seen as a finite-dimensional approximation of the Koopman operator $\mathscr{K}(t_s)$ (see \cite{Me22}). Therefore, approximations of the Koopman eigenvalues and the Koopman modes of $y$ and $\eta$ can be determined by solving an eigenvalue problem for $F$. The next theorem presents this result. 
\begin{thm}[Hankel-DMD]\label{thm:kryl}
Assume that $F$ has only the simple eigenvalues $\mu_i$, $i = 1,\ldots,n$, with associated eigenvectors $\nu_i$. Then, $\widehat{\mathcal{M}\Phi}_i = [I_3,0]H_{n}(\xi)\nu_i$, $i = 1,\ldots,n,$ is an approximation of the Koopman mode $\mathcal{M}\Phi_i$, $i \in \mathbb{N}$, of $\xi$ satisfying $\mathscr{K}(t_s)\mathcal{M}\Phi_i = \e^{\lambda_it_s}\mathcal{M}\Phi_i$ (cf. \eqref{vmod} and \eqref{xobs}) according to
\begin{equation}\label{kewpappr}
\mathscr{K}(t_s)\widehat{\mathcal{M}\Phi}_i = \e^{\hat{\lambda}_it_s}\widehat{\mathcal{M}\Phi}_i + r_{3}e\t_{n}\nu_i, \quad i \in \mathbb{N},
\end{equation}
with $r_3 = [I_3,0]r_n \in \mathbb{R}^3$ and $\hat{\lambda}_i = \ln\mu_i/t_s$. 
\end{thm}
\begin{pf}	
In view of $F\nu_i = \mu_i\nu_i$, $i = 1,\ldots,n$, the result \eqref{kewpappr} directly follows from postmultiplying \eqref{fdmd} by $v_i$ and taking only the first three rows into account, which yields $\widehat{\mathcal{M}\Phi}_i = [I_3,0]H_n(\xi)v_i$  (see  \eqref{xobs} and \eqref{Hmat}). Consider the spectral representation $\mathcal{T}_e(t_s)h = \sum_{i=1}^{\infty}\langle h,\Psi_i\rangle\e^{\lambda_it_s}\Phi_i$ (see \cite[Th. 3.2.8]{Cu20}) and \eqref{Kop}. With this, it directly follows that $\mathscr{K}(t_s)\mathcal{M}\Phi_i = \mathcal{M}\mathcal{T}_e(t_s)\Phi_i = \sum_{j=1}^{\infty}\langle\Phi_i,\Psi_j\rangle\e^{\lambda_jt_s}\mathcal{M}\Phi_j = \e^{\lambda_it_s}\mathcal{M}\Phi_i$, $i \in \mathbb{N}$, because $\langle\Phi_i,\Psi_j\rangle = \delta_{ij}$ due to the biorthormality of the sequences $\{\Psi_i,$ $i \in \mathbb{N}_0\}$ and $\{\Phi_i,$ $i \in \mathbb{N}_0\}$. As a result, $\mathcal{M}\Phi_i$ are the Koopman modes of $\xi = \mathcal{M}x_e$ w.r.t. the Koopman eigenvalues $\lambda_i$. Hence, $\widehat{\mathcal{M}\Phi}_i$ in Theorem \ref{thm:kryl} and $\hat{\lambda}_i$ are the corresponding approximations. \hfill $\Box$
\end{pf}
Note that the error $\|r_3\||e_n\t\nu_i| = \|\mathscr{K}(t_s)\widehat{\mathcal{M}\Phi}_i - \e^{\hat{\lambda}_it_s}\widehat{\mathcal{M}\Phi}_i\|$ in \eqref{kewpappr} is minimized by Lemma \ref{lem:dmd}, so that $\widehat{\mathcal{M}\Phi}_i$ and $\hat{\lambda}_i$ are also obtained in this least square sense. It is well-known that the Hankel-DMD is not numerically well-posed due to the application of the Krylov sequences for analyzing the data (see, e.g., \cite[Ch. 7]{Mau21} for a discussion). This, however, is not a serious issue for the considered setup, since it is sufficient  to only determine a few dominant modes for control. Alternatively, it is possible to directly apply the \emph{SVD-enhanced Hankel-DMD} methods presented in \cite{Ar17,Che12} to the Hankel matrix \eqref{Hmat} if a numerical stable algorithm is required. To this end, consider the real-valued \emph{economy-sized SVD} 
\begin{equation}\label{svd}
 H_{n}(\xi) = U_{n}\Sigma_{n}W\t
\end{equation}
of $H_{n}(\xi)$ in \eqref{Hmat} (see, e.g., \cite[Ch. 1]{Br22}). Introduce the matrix $H^+_{n}(\xi) = \mathscr{K}(t_s)H_{n}(\xi)$, which requires the additional snapshot $\xi(2n-1)$ (cf. \eqref{ydata} and \eqref{Hmat}). Let $\bar{v}_i$, $i = 1,\ldots,n$, be the eigenvectors of $\tilde{A} = U_{n}\t H^+_{n}(\xi)W\Sigma_{n}^{-1} \in \mathbb{R}^{n \times n}$ w.r.t. the eigenvalues $\bar{\lambda}_i$, i.e., $\tilde{A}\bar{v}_i = \bar{\lambda}_i\bar{v}_i$, $i = 1,\ldots,n$. Then, 
\begin{equation}\label{kmod22}
 H_{n}(\xi){\nu}_i = U_{n}\bar{v}_i, \quad i = 1, \ldots, n,
\end{equation}
and $\tilde{\lambda}_i = \ln\bar{\lambda}_i/t_s$ (see \cite{Ar17}).  With this, the Koopman modes and eigenvalues of Theorem \ref{thm:kryl} can directly be computed (see \cite{Che12}). The proof of this result directly follows from \cite{Sch10}. Note that $H^+_{n}(\xi) \approx AH_{n}(\xi)$ holds with $A = U_n\tilde{A}U_n\t \in \mathbb{R}^{3n \times 3n}$.
\begin{rem}\label{rem:nchoose}
The result of Theorem \ref{thm:exct} can be used to determine the sampling time $t_s$ and the number $2n$ of samples $y(k)$, $\eta(k)$ in \eqref{ydata}. For this, the Frobenius norm of the residuum $\|R_{svd}\|_F = \|\mathscr{K}(t_s)H_n(\xi) - U_n\tilde{A}U_n\t H_n(\xi)\|_F$  resulting from the SVD-enhanced Hankel-DMD  is determined as a function of $t_s$ and $n$, which directly can be obtained from data. Then, the sampling time $t_s$ and the number $n$ is chosen in given intervals such that the corresponding norm $\|R_{svd}\|_F$ is minimized. \hfill $\triangleleft$	 
\end{rem}	

Convergence of the Hankel-DMD for ODEs in the large data limit is verified in \cite{Ar17,Me22}. For parabolic systems the eigenvalues $\lambda_i$, $i \in \mathbb{N}$, grow with $|\lambda_i| \in O(i^2)$ (see, e.g., \cite{Orl17a}) so that only all unstable and the slowly decaying modes are relevant for the data. This leads to the next theorem, which presents conditions for exact Hankel-DMD, i.e., $r_n = 0$ in \eqref{fdmd} if the initial condition $x_e(0)$ can be represented by the dominant modes only, i.e., \eqref{AB} is satisfied. 
\begin{thm}[Exact Hankel-DMD]\label{thm:exct}
Let $\varphi_i[x_e]$ be the eigenfunctionals of the Koopman operator $\mathscr{K}(t)$ w.r.t. the Koopman eigenvalues $\lambda_i$ (cf. Sec. \ref{sec:koop}) and $F_d(s) \in \mathbb{C}^{3 \times q}$ be the transfer function of \eqref{plant} from $d$ to $(y,\eta)$. Assume that \eqref{AB} is fulfilled and $\rk F_d(\lambda_i) = q$, $i =1,\ldots,n_d,$ where $\lambda_i \in \sigma(S_d)$. Then, the Koopman modes $\mathcal{M}\Phi_i$ of $\xi$ are given by $ \mathcal{M}\Phi_i = [I_3,0]H_{n}(\xi)\nu_i$, $i = 1,\ldots,n,$ with the Koopman eigenvalues $\lambda_i = \ln\mu_i/t_s$ (cf. Th. \ref{thm:kryl}).
\end{thm}
\begin{pf}	
By introducing the bounded linear operator $\mathcal{A}_d = \mathcal{T}_e(t_s) : X \to X$ the Hankel matrix $H_n(\xi)$ in \eqref{Hmat} has the representation
\begin{equation}
 H_n(\xi) = \underbrace{\begin{bsmallmatrix}
 \mathcal{M}\\
 \mathcal{M}\mathcal{A}_d\\
 \vdots\\
 \mathcal{M}\mathcal{A}^{n-1}_d 
 \end{bsmallmatrix}}_{\mathcal{Q}_o}
 \begin{bmatrix}
 x_e(0) & \mathcal{A}_dx_e(0) & \ldots & \mathcal{A}^{n-1}_dx_e(0)\end{bmatrix}.
\end{equation}
By assumption one has $x_e(0) = \sum_{i=1}^n\langle x_e(0),\Psi_i\rangle\Phi_i = \underline{\Phi}_{n}\underline{\varphi}_{n}$, in which $\underline{\Phi}_{n} = [\Phi_1 \;\; \ldots \;\;\Phi_n]$ and $\underline{\varphi}_{n} = \col{\varphi_1[x_e(0)],\ldots,\varphi_n[x_e(0)]}$ are used. Then, $ H_n(\xi) = \mathcal{Q}_o [\underline{\Phi}_{n}\underline{\varphi}_{n} \ \mathcal{A}_d \underline{\Phi}_{n}\underline{\varphi}_{n} \ \ldots \  \mathcal{A}^{n-1}_d \underline{\Phi}_{n}\underline{\varphi}_{n}] = \mathcal{Q}_o\underline{\Phi}_{n} [\underline{\varphi}_{n} \ \Lambda_n\underline{\varphi}_{n} \allowbreak \ldots \ \Lambda^{n-1}_n\underline{\varphi}_{n}]$ follows in view of $\mathcal{A}_d\underline{\Phi}_{n} = \underline{\Phi}_{n}\Lambda_n$, $\Lambda_n = \diag{\lambda_1,\ldots,\lambda_n}$ and $\lambda_i \in \sigma(\mathcal{A}_e)$.  Using $\Lambda^k_n\underline{\varphi}_{n} = \diag{\varphi_i[x_e(0)]}\col{\lambda_i^k}$, $k = 0,\ldots,n-1$, gives
$H_n(\xi) = \mathcal{Q}_o\underline{\Phi}_{n}\diag{\varphi_1[x_e(0)],\ldots,\allowbreak\varphi_n[x_e(0)]}V_F$ with the \emph{Vandermonde matrix} $V_F = [1_n,\col{\lambda_i},\ldots,\col{\lambda_i^{n-1}}]$ and $1_n = \col{1,\ldots,1} \in \mathbb{R}^n$. Therein, 
\begin{equation}
Q_o = \mathcal{Q}_o\underline{\Phi}_{n}  
= \begin{bsmallmatrix}
M\\
M\Lambda_n\\
\vdots\\
M\Lambda^{n-1}_n
\end{bsmallmatrix} \in \mathbb{C}^{3n \times n}
\end{equation}
with $M = \mathcal{M}\underline{\Phi}_n \in \mathbb{C}^{3 \times n}$ holds. Since $\det\diag{\varphi_1[x_e(0)],\allowbreak\ldots,\varphi_n[x_e(0)]} \neq 0$ by assumption and $\det V_F \neq 0$ for distinct eigenvalues $\lambda_i$, $i = 1,\ldots,n$, (see \cite[Fact 7.18.5]{Bern18} and Lemma \ref{lem:exsys}), the Hankel matrix $H_n(\xi) \in \mathbb{R}^{3n \times n}$ satisfies $\rk H_n(\xi) = n$ iff $\rk Q_o = n$. Since $\Lambda_n$ has distinct eigenvalues, this holds iff $M$ has no zero columns (see \cite{Gi63}). Hence, with $M = [m_1,\ldots,m_n]$ it remains to show that 
\begin{equation}\label{mi}
 m_i = 
    \begin{bmatrix}
 	 g_4\t P_dv_i + \mathcal{C}_{z_0}\phi_i\\
 	 G_5 P_dv_i + \begin{bsmallmatrix} \mathcal{C}_0\phi_i \\ \mathcal{C}_1\phi_i
\end{bsmallmatrix} 	 
 	\end{bmatrix} \neq 0, \quad i = 1,\ldots,n,  
\end{equation}
in which $v_i$ and $\phi_i$ are determined by the eigenvectors in \eqref{vmod}. A straightforward calculation shows that
$\phi_i(z) = \phi(z,\mu_i) = (\cos(\sqrt{-\mu_i}z)  + q_0\tfrac{\sin(\sqrt{-\mu_i}z)}{\sqrt{-\mu_i}})\phi_i(0) +  r\t(z,\mu_i)P_dv_i$
with 
\begin{align}
        \mu_i &= \mu(\lambda_i) = (\lambda_i - a)/\rho,\label{muidef}\\
 r\t(z,\mu_i) \!&=\! \tfrac{\sin(\sqrt{-\mu_i}z)}{\sqrt{-\mu_i}}g_2\t \!\!-\!\! \tint_0^z\!\tfrac{\sin(\sqrt{-\mu_i}(z-\zeta))g_1\t(\zeta)}{\rho\sqrt{-\mu_i}}\d\zeta
\end{align}
and $\phi_i(0) = (r\t(1,\mu_i) + g_3\t - \text{d}_z r\t(1,\mu_i)) P_dv_i)/ h_i$. Therein,  $h_i = h(\lambda_i) = (-\sqrt{-\mu_i}-q_0/\sqrt{-\mu_i})\sin\sqrt{-\mu_i} + (q_0-q_1)\cos\sqrt{-\mu_i} \neq 0$, because $\lambda_i \notin \sigma(\mathcal{A})$, $i = 1,\ldots,n_d$. By making use of the concept of a transfer function introduced in \cite{Zw04}, i.e., using $u(t) = \e^{st}$, $s \in \mathbb{C}$, $t \geq 0$, it is easily verified that $F_d(s) = \col{g_4\t + \mathcal{H}_{z_0},~ G_5 + \col{\mathcal{H}_0 ,~ \mathcal{H}_1}}$ with $\mathcal{H}_{z} = \mathcal{C}_{z}(\cos(\sqrt{-\mu(s)}z) + q_0 (\sin(\sqrt{-\mu(s)}z) / \sqrt{-\mu(s)})) (r\t(1,\mu(s)) + g_3\t - \text{d}_z r\t(1,\allowbreak \mu(s))) / h(s)$, $z\in\{0,1\}$. Hence, $m_i = F_d(\lambda_i)P_dv_i \neq 0$, $\forall P_dv_i \neq 0$ in view of \eqref{xobs} if $ \rk F_d(\lambda_i) = q$ (see \eqref{fdmd}). For $i = n_d+1,\ldots,n$ it follows that $v_i = 0$, $\mathcal{C}_0\phi_i = \phi_i(0) \neq 0$ and $\mathcal{C}_1\phi_i = \phi_i(1) \neq 0$ so that also \eqref{mi} and thus $\rk H_n(\xi) = n$ hold. In order to verify that the eigenvalues of $F$ and $\Lambda_n$ coincide, consider $\mathscr{K}(t_s)H_n(\xi) = H_n(\xi)F$ implied by $\rk H_n(\xi) = n$ (see \eqref{fdmd}). Using $\mathcal{Q}_o\mathcal{A}_d^kx_e(0) = Q_o\Lambda_n^k\underline{\varphi}_n$ it is easy to verify that $Q_o(\Lambda^n_n + \sum_{i=0}^{n-1}f_i\Lambda_n^i)\underline{\varphi}_n = 0$ so that  $(\Lambda^n_n + \sum_{i=0}^{n-1}f_i\Lambda_n^i)\underline{\varphi}_n = 0$ follows as $\rk Q_o = n$. This directly shows that the characteristic polynomial $p(s) = s^n + \sum_{i=0}^{n-1}f_is^i$ of $F$ has the roots $\lambda_i$, $i = 1,\ldots,n$. Then, $[I_3,0]H_n(\xi)\nu_i\varphi_i[x_e(0)] = \mathcal{M}\Phi_i\varphi_i[x_e(0)]$, because $V_F\nu_i = e_i$. The latter property follows from the fact that $V_F$ is the inverse modal matrix of $F$ (see, e.g., \cite[Fact 7.18.9]{Bern18}). Hence, the equation for $\mathcal{M}\Phi_i$ in Theorem \ref{thm:exct} holds, where the factor $\varphi_i[x_e(0)]$ is absorbed in the computed Koopman mode.
\phantom{leer}\hfill $\Box$
\end{pf}
\begin{rem}	
The condition $\rk F_d(\lambda_i) = q$ in Theorem \ref{thm:exct} implies $q \leq 3$, since $F_d(s) \in \mathbb{C}^{3 \times q}$. Furthermore, no eigenmode of the disturbance model \eqref{sds} is blocked by the system so that $y$ and $\eta$ is affected by all eigenmodes of \eqref{sds}. This is necessary to obtain the Koopman eigenvalues and modes related to the disturbance state $\omega_d$ from the  output data \eqref{ydata}. Due to the boundary output $\eta$, however, all Koopman modes of the state $x$ are observable in $\eta$. 
\hfill $\triangleleft$	 
\end{rem}	

\subsection{Solution of the inverse Sturm-Liouville problem}\label{sec:para}
In what follows, it is shown that only \emph{two} Koopman modes $\mathcal{M}\Phi_{n_d+i}$, $i = 1,2$, related to the outputs $y$ and $\eta$ of  \eqref{plant}, i.e., $n \geq n_d+2$, with corresponding Koopman eigenvalues $\lambda_{n_d+1}$, $\lambda_{n_d+2}$ are sufficient to determine the parameters $\mu_i$, $q_i$, $i = 0,1$, of the related eigenvalue problem
\begin{subequations}\label{phiup}
	\begin{align}
	\phi_i''(z) &= \mu_{i}\phi_i(z), \quad z \in (0,1), i = n_d+1,n_d+2\label{parasyse1}\\
	\phi_i'(0)  &= q_0\phi_i(0), \quad \phi_i'(1)  = q_1\phi_i(1) \label{parasys2e2}
	\end{align} 	 
\end{subequations}
(cf. \eqref{phiu} and \eqref{muidef}), which amounts to solving an \emph{inverse Sturm-Liouville problem} (see, e.g., \cite{Kra20}). 
The constants $a$ and $\rho$  can be obtained from
\begin{equation}\label{modepar}
 \begin{bmatrix}
 a\\
 \rho
 \end{bmatrix}
 = 
 \begin{bmatrix}
  1 & \;\mu_{n_d+1}\\
  1 & \;\mu_{n_d+2}
 \end{bmatrix}^{-1}
  \begin{bmatrix}
 \lambda_{n_d+1}\\
 \lambda_{n_d+2}
 \end{bmatrix}
\end{equation}
(cf. \eqref{muidef}). Therein, the inverse matrix exists, since the eigenvalues of \eqref{plant} are distinct. The remaining unknown parameters $\mu_{n_d+1}$, $\mu_{n_d+2}$, $q_0$ and $q_1$ follow from the set of equations
\begin{subequations}\label{modeset}
\begin{align}
 &\mathcal{C}_0\phi_k = \phi_k(0)\\
 &\mathcal{C}_{z_0}\phi_k \!=\!(\cos(\!\sqrt{-\mu_k}z_0) \!+\!\! \tfrac{q_0}{\sqrt{-\mu_k}}\sin(\!\sqrt{-\mu_k}z_0))\mathcal{C}_0\phi_k\label{eqCZ0}\\
 &\mathcal{C}_1\phi_k = (\cos\sqrt{-\mu_k} + \tfrac{q_0}{\sqrt{-\mu_k}}\sin\sqrt{-\mu_k})\mathcal{C}_0\phi_k \label{eqC1}\\
 &(-\sqrt{-\mu_k}\sin\!\sqrt{-\mu_k} \!+\! q_0\cos\sqrt{-\mu_k})\mathcal{C}_0\phi_k\! =\! q_1 \mathcal{C}_1\phi_k\label{eqCBC1}
\end{align}	
\end{subequations}
for $k \in \{n_d+1,n_d+2\}$. Therein, \eqref{eqCZ0} is obtained from the solution of \eqref{phiup}, \eqref{eqC1} is the evaluation of the solution of \eqref{phiup} at $z=1$ and \eqref{eqCBC1} follows from evaluating the BC \eqref{parasys2e2}. If $\cos\sqrt{-\mu_k} \neq 0$ for $k \in\{n_d+1,n_d+2\}$, $q_0$ and $q_1$ can be eliminated by solving \eqref{eqCBC1} for $q_0$ and inserting the result in \eqref{eqCZ0} and \eqref{eqC1}. After equating the results for $q_1$, a transcendental equation follows for $\mu_i$, which has to be solved numerically. Note that this equation may have multiple solutions. By making use of the determined Koopman eigenvalues, however, a valid solution can be verified by solving the eigenvalue problem with the resulting parameters. Then, $q_0$ is obtained from \eqref{eqC1} and $q_1$ from \eqref{eqCBC1}. Since the modes $\mathcal{C}_{z_0}\phi_k$, $\mathcal{C}_{z_0}\phi_k$ and $\mathcal{C}_{0}\phi_k$  are directly obtained from the Koopman modes contained in $\mathcal{M}\Phi_k$, the solution of \eqref{modepar} and \eqref{modeset} can be determined from the data \eqref{ydata} using Theorem \ref{thm:kryl}.

\begin{rem}	
It is interesting to remark that the inverse Sturm-Liouville problem can also be solved for spatially varying coefficients (see \cite[Ch. 13]{Kra20}). Then, however, an infinite number of Koopman modes are required to theoretically determine the spatially varying coefficients. 	
\phantom{leer}\hfill $\triangleleft$	 
\end{rem}	
\begin{rem}\label{rem:ad}
A benefit of the proposed data-driven approach is the fact that the disturbance input vectors $g_i$, $i = 1,\ldots,4$, and $G_5$ in \eqref{plant}  have not to be determined to get the system parameters. They are also not required for the design of the regulator. This is different from identification methods based on boundary data using directly the PDE without an approximation  (see \cite{Ru08a,Kn13,Gehr16}). In addition, since only two Koopman modes with corresponding eigenvalues are required for the solution of the inverse Sturm-Liouville problem, one can choose two dominant modes. Since they are well represented in the sequential data, the latter can be determined with small errors using Hankel-DMD so that the parameters $\rho$, $a$, $q_0$ and $q_1$ are obtained accurately.
\hfill $\triangleleft$	 
\end{rem}	

\section{Nominal regulator design}\label{sec:sfr}
This section presents a data-driven solution of the robust output regulation problem posed in Section \ref{sec:probform}. For this, the parameters $\hat{\rho}$, $\hat{a}$, $\hat{q}_0$ and $\hat{q}_1$ of nominal system \eqref{plantnom} are identified. This leads to the \emph{identified system}, which is the basis for the subsequent design.

The eigenvalues $\sigma_d = \{\lambda_1,\ldots,\lambda_{n_d}\}$ of the matrix $S_d$ are obtained from the data \eqref{ydata} in Theorem \ref{thm:kryl}. Hence, any matrix $\tilde{S}_d \in \mathbb{R}^{n_d \times n_d}$ satisfying $\sigma(\tilde{S}_d) = \sigma_d$ is sufficient to describe the dynamics of the disturbance $d$. With this and \eqref{rmod} the joint signal model 
\begin{equation}\label{sigmod}
 \dot{\omega}(t) = \tilde{S}\omega(t), \quad t>0, \omega(0) = \omega_0 \in \mathbb{R}^{n_{\omega}},
\end{equation}
results with $\tilde{S} = \diag{\tilde{S}_d,S_r}$ and $n_{\omega} = n_d + n_r$ (cf. Section \ref{sec:probform}). Let $\tilde{S}_{\text{min}} \in \mathbb{R}^{\bar{n}_{\omega} \times \bar{n}_{\omega}}$, $\bar{n}_{\omega} \leq n_{\omega}$, be the \emph{cyclic part} of $\tilde{S}$.  Then, there exists a vector $v \in \mathbb{R}^{\bar{n}_{\omega}}$ such that $\det [v, \tilde{S}_{\text{min}}v, \ldots, \tilde{S}_{\text{min}}^{\bar{n}_{\omega}-1}v] \neq 0$ (see, e.g., \cite[Ch. 2.3.4]{Ack85}). This property is fulfilled if any possibly multiple eigenvalue of $\tilde{S}_{\text{min}}$ has only one linear independent eigenvector. Then, an \emph{internal model} for robust output regulation is given by
\begin{equation}\label{intmod}
 \dot{\varpi}(t) = \tilde{S}_{\text{min}}\varpi(t) + b_y(y(t) - r(t)), \;\;t > 0, 
\end{equation}
with IC $\varpi(0) = \varpi_0 \in \mathbb{R}^{\bar{n}_{\omega}}$. The vector $b_{y} \in \mathbb{R}^{\bar{n}_{\omega}}$ is chosen such that $(\tilde{S}_{\text{min}},b_{y})$ is controllable, which is always possible due to the cyclicity of $\tilde{S}_{\text{min}}$.
\begin{rem}	
In \eqref{sigmod} it is assumed that the spectrum $\sigma(S_d)$ can be determined exactly from the data. This is justified, because the disturbances are assumed to be persistently exciting and thus are well represented in the sequential data. Hence, they can be determined accurately using the Hankel-DMD. \hfill $\triangleleft$	 
\end{rem}	

The design of the robust state feedback regulator is determined to stabilize the identified system extended with \eqref{intmod} for $r = 0$. For this, the known parameters $\hat{\rho}$, $\hat{a}$, $\hat{q}_0$ and $\hat{q}_1$ are used in the design. For this, the state feedback 
\begin{equation}\label{controller}
 u = -k\t_\varpi \varpi - k_1x(1) - \tint_0^1k_x(\zeta)x(\zeta)\d\zeta
\end{equation}
is determined, which follows the lines of \cite{Deu16a,Deu20}. Consider the invertible \emph{backstepping transformation}
\begin{equation}\label{btrafo}
 \tilde{x} = x - \tint_0^z\hat{k}(z,\zeta)x(\zeta)\d\zeta = \mathcal{T}[x], 
\end{equation}

where the kernel $\hat{k}(z,\zeta)$ is the solution of the \emph{kernel equations}
\begin{subequations}\label{keqn}
	\begin{align}
&	\hat{\rho}\hat{k}_{zz}(z,\zeta) - \hat{\rho}\hat{k}_{\zeta\zeta}(z,\zeta) = (\hat{a} + \mu_c)\hat{k}(z,\zeta)\\
&	\hat{k}_\zeta(z,0) = \hat{q}_0\hat{k}(z,0), \quad \hat{k}(z,z) = \hat{q}_0-\tfrac{\hat{a} + \mu_c}{2\hat{\rho}}z
	\end{align}	 
\end{subequations}
defined on $0 \leq \zeta \leq z \leq 1$. In \cite{Sm04} it is verified that there exists a unique $C^2$-solution. Note that the kernel equations \eqref{keqn} use the known values $\hat{\rho}$, $\hat{a}$ and $\hat{q}_0$ so that their solution can be determined from the data \eqref{ydata}. The inverse transformation $x = \tilde{x} + \tint_0^z\hat{k}_I(z,\zeta)\tilde{x}(\zeta)\d\zeta = \mathcal{T}^{-1}[\tilde{x}]$ directly follows from the corresponding reciprocity relation (see \cite[Ch. 4.5]{Kr08}). A straightforward calculation shows that \eqref{btrafo} maps \eqref{plantnom} and \eqref{intmod} into
\begin{subequations}\label{plantnomcc}
	\begin{align}
	\dot{\tilde{x}} &= \rho \tilde{x}'' - \mu_c\tilde{x} + \Delta \tilde{a}(z)\tilde{x} + \Delta a_0(z)\tilde{x}(0)\nonumber\\
	&\quad  + \tint_0^z\Delta \tilde{k}(z,\zeta) \tilde{x}(\zeta)\d\zeta\label{parasysnc}\\
	\tilde{x}'(0) &= \Delta\hat{q}_0\tilde{x}(0)\\ 
	\tilde{x}'(1) &=  -k\t_\varpi \varpi - \tint_0^1(k_x(\zeta)-\hat{k}_z(1,\zeta))x(\zeta)\d\zeta \nonumber \\
	&\quad +\Delta \hat{q}_1 \tilde{x}(1) + \tint_0^1 \Delta \hat{q}_1 \hat{k}_I(1,\zeta) \tilde{x}(\zeta)\d\zeta \label{bc1}\\
	\dot{\varpi} &= \tilde{S}_{\text{min}}\varpi + b_y\mathcal{T}^{-1}[\tilde{x}](z_0)\label{ODEsub}
	\end{align}
\end{subequations}
with  the perturbations
\begin{subequations}
	\begin{align}
	\Delta\hat{a} &= a - \hat{a}, \quad \Delta\hat{\rho} = \rho - \hat{\rho}\\
	\Delta \tilde{a}(z) &= \Delta\hat{a}  + 2\Delta\hat{\rho} \,\d_z\hat{k}(z,z)\\
	\Delta\hat{q}_i &= q_i - \hat{q}_i, \quad i=0,1\\
	\Delta a_0(z) &=\rho\Delta\hat{q}_0\hat{k}(z,0)\\
	\Delta \tilde{k}(z,\zeta) &= \Delta \tilde{a}(z)\hat{k}_I(z,\zeta) + \Delta k(z,\zeta)\nonumber\\
	&\quad + \tint_{\zeta}^z\Delta k(z,\bar{\zeta})\hat{k}_I(\bar{\zeta},\zeta)\d\bar{\zeta}\\
	\Delta k(z,\zeta) &= -\Delta\hat{a} \hat{k}(z,\zeta) + \Delta\hat{\rho}\hat{k}_{zz}(z,\zeta)\nonumber\\
	&\quad  - \Delta\hat{\rho}\hat{k}_{\zeta\zeta}(z,\zeta),                                     
	\end{align}	 
\end{subequations}
where $k_1 = \hat{q}_1-\hat{k}(1,1)$. In the final step, the ODE subsystem \eqref{ODEsub} is decoupled from the PDE subsystem by making use of the \emph{decoupling transformation}
\begin{equation}\label{detraf}
e = \varpi  - \tint_0^1\hat{\tilde{q}}(\zeta)\tilde{x}(\zeta)\d\zeta,
\end{equation}
in which $\hat{\tilde{q}}(z)$ is the solution of the \emph{decoupling equations}
\begin{subequations}\label{decupeq}
	\begin{align}
	\tilde{S}_{\text{min}}\hat{\tilde{q}}(z) -\hat{\rho}\hat{\tilde{q}}''(z) + \mu_c\hat{\tilde{q}}(z) &= -b_y\tilde{c}(z)\\
	\hat{\tilde{q}}'(0) &= \hat{\tilde{q}}'(1) = 0
	\end{align}	 
\end{subequations}
on $z \in (0,1)$ with $\tilde{c}(z) = \delta(z-z_0) + \hat{k}_I(z_0,z)h(z)$ and $h(z) = 1$, $z \in [0,z_0]$, $h(z) = 0$, $z \in (z_0,1]$. Note that all parameters in \eqref{decupeq} are obtained from the data \eqref{ydata}. A direct calculation shows that \eqref{decupeq} have a unique piecewise $C^2$-solution if $\mu_c \in \mathbb{R}$ is such that $\dot{w} = \hat{\rho}w'' + \mu_c w$, $w'(0) = w'(1) = 0$, is exponentially stable (see also \cite{Deu16a,Deu20}). It can be verified that 
\begin{equation}\label{eorig}
e = \varpi  - \tint_0^1\hat{q}(\zeta)x(\zeta)\d\zeta
\end{equation}
with $\hat{q}(z) = \hat{\tilde{q}}(z) - \tint_z^1\hat{\tilde{q}}(\bar{\zeta})\hat{k}(\bar{\zeta},z)\d\bar{\zeta}$ holds (see \cite{Deu16a,Deu20}). Then, inserting \eqref{eorig} in \eqref{bc1} shows that $k_x(z) = -\hat{k}_z(1,z) - k\t_\varpi\hat{q}(z)$, which leads to the BC \eqref{bce}. By applying the decoupling transformation \eqref{detraf} to \eqref{plantnomcc} the \emph{final target system}
\begin{subequations}\label{plantnomc}
	\begin{align}
	\dot{e} &= (\tilde{S}_{\text{min}} + bk\t_\varpi)e + \Delta be + \Delta g_0\tilde{x}(0)\nonumber\\
	& \quad   + \tint_0^1\Delta g(\zeta)\tilde{x}(\zeta)\d\zeta + \Delta g_1 \tilde{x}(1)\label{esys}\\ 
	\dot{\tilde{x}} &= \rho \tilde{x}'' - \mu_c\tilde{x} + \Delta \tilde{a}(z){x} + \Delta a_0(z)\tilde{x}(0)\nonumber\\
	&\quad  + \tint_0^z\Delta\tilde{k}(z,\zeta) \tilde{x}(\zeta)\d\zeta\label{parasysncf}\\
	\tilde{x}'(0) &= \Delta\hat{q}_0\tilde{x}(0)\\ 
	\tilde{x}'(1) &= -k\t_\varpi e + \Delta \hat{q}_1 \tilde{x}(1)\nonumber\\
	&\quad  + \tint_0^1 \Delta \hat{q}_1 \hat{k}_I(1,\zeta) \tilde{x}(\zeta)\d\zeta\label{bce}
	\end{align}
\end{subequations}
with $b = \hat{\rho}\hat{\tilde{q}}(1)$ and the perturbations $\Delta b = \hat{\tilde{q}}(1)\Delta\hat{\rho} k\t_\varpi$, $\Delta g_0 = -\!\!\tint_0^1\hat{\tilde{q}}(\zeta)\Delta a_0(\zeta)\d\zeta + \Delta\hat{q}_0\rho\hat{\tilde{q}}(0)$, $\Delta g_1 \!=\!  - \Delta \hat{q}_1\rho\hat{\tilde{q}}(1)$,
$\Delta g(z) = -\Delta\tilde{a}(z)\hat{\tilde{q}}(z) - \tint_{z}^1\hat{\tilde{q}}(\bar{\zeta})\Delta k(\bar{\zeta},z)\d\bar{\zeta}	- \Delta\hat{\rho}\hat{\tilde{q}}''(z) -\Delta g_1 \hat{k}_I(1,z)$ result, in which $\hat{\tilde{q}}''(z)$ is piecewise defined. Therein, a feedback gain $k\t_\varpi$ ensuring a Hurwitz matrix $\tilde{S}_{\text{min}} + bk\t_\varpi$ exists, i.e., $(\tilde{S}_{\text{min}},b)$ is controllable if the transfer function from $\hat{u}$ to $\hat{y}$ of the system
$\dot{\hat{x}}   = \hat{\rho}\hat{x}'' + \hat{a}\hat{x}$, $\hat{x}'(0) = \hat{q}_0\hat{x}(0)$, $\hat{x}'(1) = \hat{q}_1\hat{x}(1)+\hat{u}$ and 
$\hat{y} = \hat{x}(z_0)$ has no transmission zeros being an element of $\sigma(\tilde{S}_{\text{min}})$ (see \cite{Deu16a,Deu20} for a proof). Note that this can also be verified in the design as all parameter of this system follows from the given data \eqref{ydata}.

In \cite{Deu16a,Deu20} it is verified that the state feedback regulator designed for \eqref{plantnom} using the nominal parameters $\rho$, $a$, $q_0$ and $q_1$ is exponentially stable in the norm $(\|\cdot\|^2_{\mathbb{C}^{\bar{n}_\omega}} + \|\cdot\|^2_{L_2})^{1/2}$ with the decay rate $\alpha_c = \allowbreak\min(\operatorname{Re}_{\lambda \in \sigma(\tilde{S}_{\text{min}} + bk_{\bar{\omega}}\t)}\lambda,\mu_c)$. Since the parameters $\hat{\rho}$, $\hat{a}$, $\hat{q}_0$ and $\hat{q}_1$ used for the design can be determined with small errors (see Remark \ref{rem:ad}) it is sufficient to verify well-posedness and exponential stability of the corresponding closed-loop system for this case. This is  the result of the next theorem, which is proved in the Appendix.
\begin{thm}[Nominal closed-loop stability]\label{thm:stabnom}
Assume that $\tilde{S}_{\text{min}} + bk\t_\varpi$  is a Hurwitz matrix and $\mu_c > 0$. Then, the nominal closed-loop system \eqref{plantnom}, \eqref{intmod} and \eqref{controller} is exponentially stable in the norm 	$(\|\cdot\|^2_{\mathbb{C}^{\bar{n}_\omega}} + \|\cdot\|^2_{L_2})^{1/2}$ for sufficiently small errors $\Delta\hat{\rho} = \rho - \hat{\rho}$, $\Delta\hat{a} = a- \hat{a}$, $\Delta\hat{q}_0 = q_0 - \hat{q}_0$ and $\Delta \hat{q}_1 = q_1 - \hat{q}_1$.
\end{thm}

\section{Robust output regulation}\label{sec:robreg}
As the designed controller \eqref{intmod}, \eqref{controller} stabilizes the nominal closed-loop system in the presence of errors $\Delta\hat{\rho}$, $\Delta \hat{a}$, $\Delta \hat{q}_0$ and $\Delta \hat{q}_1$ resulting from the Hankel-DMD, robust output regulation \eqref{outreg} is ensured. The resulting regulator, however, can also deal with additional model uncertainties. In particular, by assuming that $\Delta\rho(z)$, $\Delta a(z)$, $\Delta q_0$ and $\Delta q_1$ both contain the Hankel-DMD errors and the model uncertainty, the corresponding uncertain closed-loop system should at least remain \emph{strongly asymptotically stable} (see, e.g., \cite{Hu93}), i.e., it is the generator of a strongly asymptotically stable $C_0$-semigroup $\mathcal{T}_c(t)$, $t \geq 0$, satisfying $\lim_{t\to\infty} \|\mathcal{T}_c(t) x_c(0)\| = 0$, for all ICs $x_c(0) = \col{\varpi(0),x(0)}$ compatible with BCs. The next theorem shows that this implies robust output regulation for the system \eqref{plant} in the presence of non-destabilizing model uncertainties. 

\begin{thm}[Robust output regulation]\label{thm:robreg}\hfill
Assume that the nominal system parameters $(\rho,$ $a,$ $q_0,$ $q_1)$ are perturbed to $(\rho + \Delta\rho(z),$ $a + \Delta a(z),$ $q_0 + \Delta q_0,$ $q_1 + \Delta q_1)$ such that the resulting closed-loop system is strongly asymptotically stable. Then, the controller \eqref{intmod}, \eqref{controller} achieves output regulation, i.e., $\lim_{t \to \infty}e_y(t) = 0$, for any  disturbance input locations characterized by $g_i$, $i = 1,\ldots,4$, $G_5$ as well as any $P_d$ and $p\t_r$ in \eqref{dmod}, \eqref{rmod}.  
\end{thm}
The proof of this theorem can be directly inferred from \cite{Deu16a,Deu20}. Note that the closed-loop system will at least be strongly asymptotically stable for sufficiently small model uncertainties.

\section{Example}\label{sec:ex}
To illustrate the result of this paper, an unstable parabolic system is considered. The nominal parameters of the plant \eqref{plant} are $\rho = 1.5$, $a = 8$, $q_0 = 2.5$ and $q_1 = -2$. The largest eigenvalue of the system is 3.196. A disturbance $d(t) \in \mathbb{R}$ acts in-domain as well as at the boundaries, which is described by $g_1(z) = 3$, $g_2 = g_3 = 1$, $g_4 = 0$ and $G_5 = 0$. The output to be controlled $y(t) \in \mathbb{R}$ is defined at $z_0 = 0.5$.
\begin{figure}[!t]
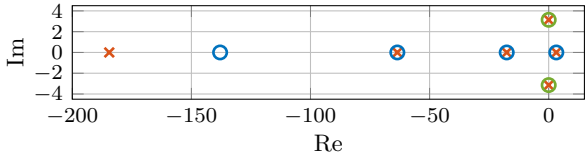

\centering
\setlength\figureheight{0.07\textwidth}
\setlength\figurewidth{0.4\textwidth}
\include{KEigenvalues}
\vspace*{-5mm}
\caption{Eigenvalues of the nominal system (\matlabBlueCirc), the disturbance model (\matlabGreenCirc) and the eigenvalues resulting from the Hankel-DMD (\matlabOrangeCross).}
\label{fig:keigvalues}
\end{figure}
First, the unknown parameters $\rho$, $a$, $q_0$, $q_1$ and the eigenvalues of the disturbance model must be identified using the Hankel-DMD, described in Subsection \ref{subsec:hankeldmd}, with the output data \eqref{ydata} generated by a simulation. The disturbance $d(t)=\sin(\omega_d t)$, $t \geq -0.1$, $\omega_d = \pi$, persistently acts on the system and the IC follows from applying a rectangular impulse $u(t) = 10 (s(t+0.1)-s(t))$ and $x(z,-0.1)=0$, where $s(t)$ is the step function. The dominant modes are then well contained in the output data \eqref{ydata} for $t\geq 0$. Afterwards, the SVD-enhanced Hankel-DMD is used, because of numerically well-posedness. By calculating the norm of the residuum $R_{svd}$ over different sampling times $t_s$ and number of sampling values $n$ the local minimum $\|R_{svd}\|_F = 6.862 \cdot 10^{-12}$ is found leading to $t_s = 0.104$ and $n = 6$ (see Remark \ref{rem:nchoose}). It is verified that the resulting eigenvalues fit to the eigenvalues obtained from the inverse Sturm-Liouville problem with the identified parameters.
The plot in Fig. \ref{fig:keigvalues} shows the identified eigenvalues and the exact eigenvalues of the PDE-ODE system \eqref{plant} and \eqref{dmod}. Obviously, the dominant plant and all disturbance model eigenvalues are captured very well by the Hankel-DMD. Using the two slowest plant eigenvalues and the corresponding Koopman modes, the approximated parameters $\hat{\rho} = 1.510$, $\hat{a} = 8.020$, $\hat{q}_0 = 2.491$, $\hat{q}_1 = -1.992$ and $\hat{\omega}_d = 3.1416$ are obtained by numerically solving an inverse Sturm-Liouville problem with MATLAB.
Afterwards the regulator design can be carried out as described in Section \ref{sec:sfr} to ensure the tracking of the ramp reference signal $r(t) = q_{r0} + q_{r1} t$, $t >0$, $q_{ri} \in \mathbb{R}$, $i = 0,1,$ which can be periodically reset to achieve sawtooth waves. The corresponding internal model is determined by
\begin{equation}
\tilde{S} = \tilde{S}_{\text{min}} = \text{diag}(\begin{bsmallmatrix}0&1 \\ -\hat{\omega}_d^2&0\end{bsmallmatrix}, \begin{bsmallmatrix}0&1 \\ 0&0 \end{bsmallmatrix})
\end{equation}
and $b_y = [1\ 0\ 0\ 1]\t$, so that the pair $(\tilde{S}_{\text{min}},b_{y})$ is controllable. With this and solving the decoupling equations \eqref{decupeq} in MATLAB, the feedback gain $k\t_\varpi$ can be computed to place the eigenvalues of $\tilde{S}_{\text{min}} + bk\t_\varpi$ at $-4.5 \pm \text{j}\pi$, $-4$ and $-5$.
The design parameter of the backstepping-based state feedback controller is chosen to $\mu_c = 5$ and the kernel equations \eqref{keqn} are solved using the method of successive approximations (see \cite{Sm04}).
\begin{figure}[!t]
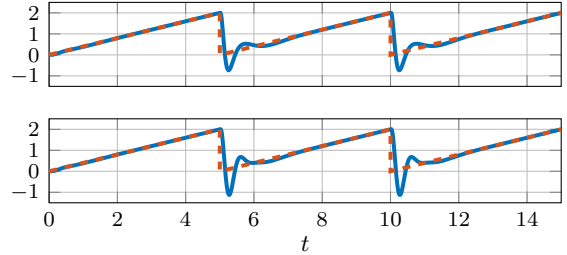

\centering
\setlength\figureheight{0.15\textwidth}
\setlength\figurewidth{0.4\textwidth}
\include{refBehavior}
\vspace*{-5mm}
\caption{Closed-loop reference behavior of $y(t)$ (\matlabBlueLine) for $r(t) = \tfrac{2}{5}\mod(t,5)$ (\matlabOrangeLine) and $d(t) \equiv 0$. Upper plot shows the nominal case and lower plot with model uncertainty.}
\label{fig:refbehavior}
\end{figure}
\begin{figure}[!t]
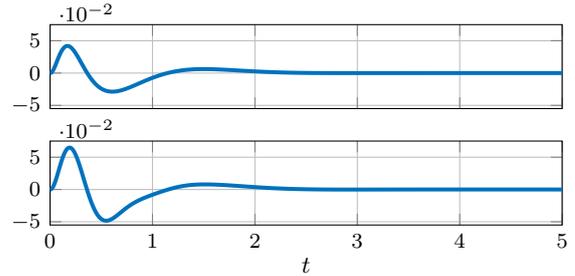

\centering
\setlength\figureheight{0.15\textwidth}
\setlength\figurewidth{0.4\textwidth}
\include{distBehavior}
\vspace*{-5mm}
\caption{Closed-loop disturbance behavior of $y(t)$ (\matlabBlueLine) for $d(t) = \sin(\pi t)$ and $r(t) \equiv 0$. Upper plot shows the nominal case and lower plot with model uncertainty.}
\label{fig:distbehavior}
\end{figure}
To investigate the robustness of the designed regulator, two simulations are carried out with vanishing ICs for the controller and the plant. On the one hand, the nominal system and, on the other, model perturbations of 70 percent of all nominal parameters are considered.
The controller achieves asymptotic tracking of the reference signal in the nominal case (see Fig. \ref{fig:refbehavior}, upper plot) as well as in the presence of model uncertainty (see Fig. \ref{fig:refbehavior}, lower plot). In Fig. \ref{fig:distbehavior}, the disturbance rejection achieved by the robust regulator for the nominal case and with model uncertainty is depicted. This verifies data-driven robust output regulation.

\section{Concluding remarks}
It is straightforward to extend the considered state feedback regulator by a backstepping observer using the available data. Further research considers systems with spatially varying coefficients and the extension to second-order systems like beams and wave equations. For applications it is of interest to investigate the influence of measurement noise in the data-driven approach.

\appendix
\section{Proof of Theorem \ref{thm:stabnom}}
The proof proceeds in several steps. In the first step, the operator $\mathcal{A}_1h = \rho h'' - \mu_c h$ with $D(\mathcal{A}_1) = \{ h \in H^2(0,1)\,|\, h'(0) = \Delta\hat{q}_0h(0),h'(1) = \Delta\hat{q}_1h(1) + \tint_0^1\Delta\hat{q}_1\hat{k}_I(1,\zeta)h(\zeta)\d\zeta\}$ is considered.  A straightforward calculation shows that the adjoint operator reads $\mathcal{A}^*_1h = \rho h'' - \mu_c h + \rho \Delta\hat{q}_1\hat{k}_I(1,\cdot)h(1)$ with $D(\mathcal{A}^*_0) = \{ h \in H^2(0,1)\,|\, h'(0) = \Delta\hat{q}_0h(0),h'(1) = \Delta\hat{q}_1h(1)\}$. Introduce the operators $\mathcal{A}_0h = \rho h'' -\mu_ch$, $D(\mathcal{A}_0) = \{h \in H^2(0,1)\;|\;h'(0) = \Delta\hat{q}_0h(0), h'(1) = \Delta\hat{q}_1h(1)\}$ and $\mathcal{B}_0h = \rho\Delta\hat{q}_1 \hat{k}_I(1,\cdot)h(1)$, $D(\mathcal{B}_0) = D(\mathcal{A}_0)$ so that $\mathcal{A}_1^*  = \mathcal{A}_0 + \mathcal{B}_0$. The operator $\mathcal{A}_0$ is the generator of an exponentially stable $C_0$-semigroup $\mathcal{T}_{\mathcal{A}_0}(t)$, $t \geq 0$, for $\mu_c > 0$, because $\mathcal{A}_0$ is Riesz-spectral and generates for $\Delta\hat{q}_0 = \Delta\hat{q}_1 = 0$ an exponentially stable $C_0$-semigroup, i.e., $\|\mathcal{T}_{\mathcal{A}_0}(t)\| \leq M_0\e^{-\alpha_0t}$, $t \geq 0$, $\alpha_0 > 0$, $M_0 \geq 1$, holds (see, e.g., \cite[Th. 3.2.8]{Cu20}). This result follows from the fact that the eigenvalues vary continuously w.r.t. the system parameters so that for small $\Delta\hat{q}_0$,  $\Delta\hat{q}_1$ the $C_0$-semigroup $\mathcal{T}_{\mathcal{A}_0}(t)$ remains exponentially stable (see, e.g., \cite{Kon96}). In \cite[App. A.1]{Ker21} it is verified that $\mathcal{B}_0$ is \emph{$\mathcal{A}_0$-bounded}, i.e., there exist positive constants $a_0$ and $b_0$ such that $|\mathcal{B}_0h| \leq a_0\|\mathcal{A}_0h\| + b_0\|h\|$, $h \in D(\mathcal{A}_0)$, holds. Then, with $\mathcal{T}_{\mathcal{A}_0}(t_0)h \in D(\mathcal{A}_0)$, $t_0 > 0$, the result $|\mathcal{B}_0\mathcal{T}_{\mathcal{A}_0}(t_0)h| \leq a_0\|\mathcal{A}_0\mathcal{T}_{\mathcal{A}_0}(t_0)h\| + b_0\|\mathcal{T}_{\mathcal{A}_0}(t_0)h\| \leq c \|h\|$ for a $c > 0$ is easily verified, because $\mathcal{T}_{\mathcal{A}_0}(t)$, $t > 0$, is an analytic $C_0$-semigroup implying  $\mathcal{A}_0\mathcal{T}_{\mathcal{A}_0}(t)$ being a bounded linear operator (see \cite[Ex. 2.23]{Cu20}). Hence,  $\mathcal{B}_0\mathcal{T}_{\mathcal{A}_0}(t_0)$ is also a bounded linear operator. Consequently, $|\mathcal{B}_0\mathcal{T}_{\mathcal{A}_0}(t)h| \leq \|\mathcal{B}_0\mathcal{T}_{\mathcal{A}_0}(t_0)\mathcal{T}_{\mathcal{A}_0}(t-t_0)h\| \leq \|\mathcal{B}_0\mathcal{T}_{\mathcal{A}_0}(t_0)\|\|\mathcal{T}_{\mathcal{A}_0}(t-t_0)h\| \leq c_0M_0\e^{-\alpha_0(t - t_0)}\|h\|$, $h \in D(\mathcal{A}_0)$, $t > t_0 > 0$, with $c_0 = \|\mathcal{B}_0\mathcal{T}_{\mathcal{A}_0}(t_0)\|$ readily follows. With this, one obtains 
$L^2_{\mathcal{B}_0} = \sup_{\|h\|=1}\tint_0^\infty\|\mathcal{B}_0\mathcal{T}_{\mathcal{A}_0}(t)h\|^2\d t \leq c_0M_0\tint_0^\infty\e^{-2\alpha_0(t - t_0)}\d t$ for $h \in D(\mathcal{A}_0)$. The latter integral is finite and $\Delta\hat{q}_1$ is arbitrarily and also $c_0$ is small. Hence, $L_{\mathcal{B}_0} < 1/(2A_0)$ with $A_0^2 = \sup_{\|h\|=1}\tint_0^\infty\|\mathcal{T}_{\mathcal{A}_0}(t)h\|^2\d t < \infty$, $h \in D(\mathcal{A}_0)$, holds. Since the evaluation operator $\mathcal{C}_1h = h(1)$, $D(\mathcal{C}_1) = D(\mathcal{B}_0)$, has finite range the operator $\mathcal{B}_0$ is also $\mathcal{A}_0$-compact and thus has $\mathcal{A}_0$-bound zero (see \cite[Lem. III/2.16]{Eng00}). With \cite[Th. III/2.10]{Eng00} and \cite{Pan91} this shows that $\mathcal{A}_1^* = \mathcal{A}_0 + \mathcal{B}_0$ is the infinitesimal generator of an exponentially stable and analytic $C_0$-semigroup. Then, with the same reasoning as in \cite[Th. 8]{Deu19} one can verify that $\mathcal{A}_1$ has the same properties. In the next step, consider the operator $\mathcal{A}_2 = \mathcal{A}_1 + \mathcal{B}_1$ with $\mathcal{B}_1h = \Delta\hat{a}_1h(0)$, $D(\mathcal{B}_1) = D(\mathcal{A}_1)$. Using the same reasoning it is readily verified that $\mathcal{A}_2$ is the infinitesimal generator of an exponentially stable and analytic $C_0$-semigroup. With these preparations, the system 
\begin{subequations}\label{plantnomccpp}
	\begin{align}
	\dot{e} &= (\tilde{S}_{\text{min}} + bk_{\varpi}\t)e + \Delta g_0\tilde{x}(0) + \Delta g_1 \tilde{x}(1)\\
	\dot{\tilde{x}} &= \rho \tilde{x}'' - \mu_c\tilde{x} + \Delta a_0(z)\tilde{x}(0)\\
	\tilde{x}'(0) &= \Delta\hat{q}_0\tilde{x}(0)\\ 
	\tilde{x}'(1) &= \Delta\hat{q}_1\tilde{x}(1) \!+\! \tint_0^1\Delta\hat{q}_1\hat{k}_I(1,\zeta)h(\zeta)\d\zeta \!-\! k\t_\varpi e
	\end{align}
\end{subequations}
can be investigated. By making use of the reasoning in \cite{Deu16a,Deu20} and the previous results it is straightforward to verify that \eqref{plantnomccpp} is also exponentially stable and generates an analytic $C_0$-semigroup for $\Delta g_0 =  \Delta g_1 = 0$. Consequently, the exponential stability of the perturbed system \eqref{plantnomccpp} follows from the same reasoning as for $\mathcal{A}_1$ and $\mathcal{A}_2$.  Finally, the remaining perturbations in \eqref{plantnomc} are all bounded and are arbitrarily small for sufficiently small errors $\Delta\hat{\rho}$, $\Delta\hat{a}$, $\Delta\hat{q}_0$ and $\Delta\hat{q}_1$ so that also \eqref{plantnomc} is exponentially stable (see, e.g., \cite[Th. 5.3.1]{Cu20}). Then, the bounded invertibility of the transformations \eqref{btrafo} and \eqref{detraf} also verify exponential stability of the nominal closed-loop system \eqref{plantnom}, \eqref{intmod} and \eqref{controller}.

\bibliographystyle{plain}        
\bibliography{mybib}           




\end{document}

%% file: KEigenvalues.tex
%
%
\definecolor{mycolor1}{rgb}{0.85000,0.32500,0.09800}%
\definecolor{mycolor2}{rgb}{0.00000,0.44700,0.74100}%
\definecolor{mycolor3}{rgb}{0.4660 0.6740 0.1880}
\begin{tikzpicture}

\begin{axis}[%
width=0.951\figurewidth,
height=\figureheight,
at={(0\figurewidth,0\figureheight)},
scale only axis,
xmin=-200,
xmax=15,
xlabel style={font=\color{white!15!black}},
xlabel={Re},
ymin=-4.5,
ymax=4.5,
ylabel style={font=\color{white!15!black}},
ylabel={Im},
axis background/.style={fill=white},
title style={font=\bfseries},
xmajorgrids,
ymajorgrids,
ylabel near ticks,
xlabel near ticks,
title style={font=\normalsize},
xlabel style={font=\footnotesize},
xticklabel style ={/pgf/number format/fixed, /pgf/number format/precision=3},
ylabel style={font=\footnotesize},
yticklabel style ={/pgf/number format/fixed, /pgf/number format/precision=3},
xlabel shift={-3pt},
ylabel shift={-3pt},
zlabel shift={-5pt},
legend style={font=\scriptsize},
ticklabel style={font=\scriptsize}
]
\addplot [color=mycolor1, line width=1.0pt, only marks, mark size=2.5pt, mark=x, mark options={solid, mycolor1}, forget plot]
  table[row sep=crcr]{%
-1.49213974509621e-11	3.14159265359265\\
-1.49213974509621e-11	-3.14159265359265\\
3.19572255588335	0\\
-17.6236715868355	0\\
-63.5431067536855	0\\
-184.487518924674	0\\
};
\addplot [color=mycolor2, line width=1.0pt, only marks, mark size=2.5pt, mark=o, mark options={solid, mycolor2}, forget plot]
  table[row sep=crcr]{%
3.19572255616973	0\\
-17.623671722137	0\\
-63.5432737240054	0\\
-137.942529524427	0\\
};

\addplot [color=mycolor3, line width=1.0pt, only marks, mark size=2.5pt, mark=o, mark options={solid, mycolor3}, forget plot]
  table[row sep=crcr]{%
0	3.14159265358978\\
0	-3.14159265358978\\
};
\end{axis}
\end{tikzpicture}%

%% file: refBehavior.tex
%
%
\definecolor{mycolor1}{rgb}{0.00000,0.44700,0.74100}%
\definecolor{mycolor2}{rgb}{0.85000,0.32500,0.09800}%
\begin{tikzpicture}

\begin{axis}[%
width=0.951\figurewidth,
height=0.419\figureheight,
at={(0\figurewidth,0.581\figureheight)},
scale only axis,
xmin=0,
xmax=15,
ymin=-1.5,
ymax=2.5,
ylabel style={font=\color{white!15!black}},
xticklabels={\empty},
axis background/.style={fill=white},
xmajorgrids,
ymajorgrids,
ylabel near ticks,
xlabel near ticks,
title style={font=\normalsize},
xlabel style={font=\footnotesize},
xticklabel style ={/pgf/number format/fixed, /pgf/number format/precision=3},
ylabel style={font=\footnotesize},
yticklabel style ={/pgf/number format/fixed, /pgf/number format/precision=3},
ylabel shift={-5pt},
zlabel shift={-5pt},
legend style={font=\scriptsize},
ticklabel style={font=\scriptsize}
]
\addplot [color=mycolor1, line width=1.5pt, forget plot]
  table[row sep=crcr]{%
0	0\\
0.059003933595573	0.00110480963842008\\
0.0850056670444701	0.00421341130737218\\
0.110007333822255	0.00969291121953653\\
0.136009067271152	0.0179844327241927\\
0.16401093406227	0.0294809819547588\\
0.198013200880059	0.0461215165951749\\
0.24801653443563	0.0735647393278711\\
0.349023268217881	0.129395894250406\\
0.404026935129009	0.156773139678132\\
0.460030668711248	0.18198970998148\\
0.523034868991266	0.207701773349145\\
0.603040202680178	0.237661451528323\\
0.768051203413561	0.296191108643672\\
0.911060737382492	0.348236006153948\\
1.03806920461364	0.396969467185627\\
1.17807853856924	0.453252473827485\\
1.36309087272485	0.530316068500792\\
1.95913060870725	0.780206698115999\\
2.26815121008067	0.906119637890626\\
2.75818387892526	1.10292098099505\\
5.01233415561038	2.00413733609569\\
5.01633442229482	2.00326289461022\\
5.01933462230815	2.00034073118966\\
5.0233348889926	1.99227442515545\\
5.02733515567705	1.9785204350612\\
5.0323354890326	1.95245544269219\\
5.03833588905927	1.9077178211729\\
5.04533635575705	1.8375637765688\\
5.05433695579705	1.72193591464958\\
5.06533768917928	1.54923806198718\\
5.08133875591706	1.25706803074443\\
5.1373424894993	0.19822179533999\\
5.15334355623708	-0.0489422474972763\\
5.16834455630375	-0.245659223621105\\
5.1813454230282	-0.387502504742846\\
5.19334622308154	-0.494934973474409\\
5.20434695646376	-0.574220422188862\\
5.21434762317488	-0.6311167092113\\
5.22334822321488	-0.670653940478347\\
5.23134875658377	-0.69707657647294\\
5.23934928995266	-0.715813407620308\\
5.24634975665044	-0.726313613486026\\
5.25235015667711	-0.731218796207449\\
5.25835055670378	-0.732582737220776\\
5.26435095673045	-0.73063129481978\\
5.27035135675712	-0.725588558187976\\
5.2773518234549	-0.716093062732396\\
5.28535235682379	-0.700898054678621\\
5.2953530235349	-0.676117095527994\\
5.30635375691713	-0.64246576993459\\
5.32035469031269	-0.591724121229118\\
5.33835589039269	-0.51672147678542\\
5.36335755717048	-0.401310839985198\\
5.43836255750383	-0.0485487217365286\\
5.46336422428162	0.0553608216942791\\
5.48636575771718	0.141382874705911\\
5.50736715781052	0.21134981041067\\
5.52736849123275	0.270283745937776\\
5.54636975798387	0.319448172280071\\
5.56537102473498	0.362230019501116\\
5.58337222481499	0.397189639326148\\
5.60137342489499	0.427080924210918\\
5.61837455830389	0.450993570438069\\
5.63537569171278	0.471042196277402\\
5.65237682512167	0.487552462011635\\
5.66937795853057	0.500844968413915\\
5.68637909193946	0.51123063211549\\
5.70438029201947	0.519389793224672\\
5.72238149209947	0.524959285314399\\
5.74138275885059	0.528376167671517\\
5.76238415894393	0.529617356067757\\
5.78538569237949	0.528410379995091\\
5.81038735915728	0.5246291646015\\
5.83938929261951	0.517763741406528\\
5.87539169277952	0.506662568874585\\
5.92939529301954	0.487211524504584\\
6.01540102673512	0.456316884735367\\
6.06040402693513	0.442735181371495\\
6.10040669377959	0.433040351759612\\
6.13840922728182	0.426240917003435\\
6.17441162744183	0.422135799158379\\
6.21041402760184	0.420381035724665\\
6.24641642776185	0.420987019372674\\
6.28241882792186	0.423911463851654\\
6.31942129475298	0.42924390040444\\
6.35842389492633	0.437277388336051\\
6.3994266284419	0.448195272994901\\
6.44242949529969	0.462110670216788\\
6.48943262884192	0.479863596993594\\
6.5404360290686	0.501684640454899\\
6.59843989599307	0.529118543627932\\
6.66744449629975	0.564463628140748\\
6.75745049669978	0.613357381730795\\
7.09047269817988	0.796662850184944\\
7.19547969864658	0.850499658903296\\
7.30048669911328	0.901720793869162\\
7.41049403293553	0.952777438517449\\
7.53150210014001	1.00635268258216\\
7.67351156743783	1.0666239582609\\
7.85852390159344	1.1424757682225\\
8.17554503633576	1.26953615305422\\
9.23061537435829	1.69169287365206\\
10.0136675778385	2.0041535055838\\
10.017667844523	2.00253531433014\\
10.0206680445363	1.99879817436905\\
10.0246683112207	1.98938895444412\\
10.0296686445763	1.9692987754597\\
10.0346689779319	1.93916255826717\\
10.0406693779585	1.88958010180526\\
10.0486699113274	1.80198144675189\\
10.0586705780385	1.66263256352387\\
10.071671444763	1.44404229446251\\
10.0916727781852	1.06152224940968\\
10.1276751783452	0.370285000256464\\
10.1456763784252	0.0700291856225892\\
10.1606773784919	-0.145112872322064\\
10.1746783118875	-0.314450998618254\\
10.1876791786119	-0.444009372724148\\
10.1996799786652	-0.540410821270608\\
10.2106807120475	-0.610046583712892\\
10.2206813787586	-0.658666049118221\\
10.2296819787986	-0.691216920097764\\
10.2376825121675	-0.711826747937771\\
10.2446829788653	-0.723857727490248\\
10.2506833788919	-0.729993556292101\\
10.2556837122475	-0.73233439794069\\
10.2616841122742	-0.732012433465234\\
10.2676845123008	-0.728487434212349\\
10.2736849123275	-0.72198213532756\\
10.2806853790253	-0.710918036135853\\
10.2896859790653	-0.691716581510551\\
10.2996866457764	-0.664601121422471\\
10.3116874458297	-0.625266054918439\\
10.3266884458964	-0.567770204657544\\
10.3456897126475	-0.485401546299189\\
10.3756917127809	-0.343622351633442\\
10.4286952463498	-0.092809532462363\\
10.4556970464698	0.0232272617283495\\
10.4786985799053	0.112553769033068\\
10.5007000466698	0.188945920453149\\
10.520701380092	0.250502079927454\\
10.5397026468431	0.302089969314359\\
10.5587039135942	0.347186479529269\\
10.5767051136742	0.384212611376498\\
10.5947063137542	0.416031771975087\\
10.6117074471631	0.441630142581376\\
10.628708580572	0.463230341987815\\
10.6457097139809	0.481158380627384\\
10.6627108473898	0.495737592880037\\
10.6797119807987	0.507283347189931\\
10.6967131142076	0.516099275701114\\
10.7147143142876	0.522779672344281\\
10.7337155810387	0.527220439932874\\
10.753716914461	0.529381603454837\\
10.7757183812254	0.529229747097567\\
10.7997199813321	0.526569819387506\\
10.8267217814521	0.52111910884835\\
10.8587239149277	0.512182553049861\\
10.9017267817855	0.497538470966164\\
11.0377358490566	0.449358680910027\\
11.0797386492433	0.437821497642242\\
11.1187412494166	0.429510632843471\\
11.1557437162478	0.424010710110101\\
11.1917461164078	0.421010521794244\\
11.2277485165678	0.420372861702207\\
11.2637509167278	0.422081879345161\\
11.3007533835589	0.426216677079783\\
11.3387559170611	0.432857345243459\\
11.3787585839056	0.442299302419348\\
11.4207613840923	0.454692285941626\\
11.4657643842923	0.470488572087794\\
11.5147676511767	0.490256474397315\\
11.5687712514168	0.514620533924576\\
11.6317754516968	0.545715452439753\\
11.7087805853724	0.586450720029307\\
11.8257883858924	0.651316699530206\\
12.0218014534302	0.759905429753015\\
12.1318087872525	0.817996430242946\\
12.2358157210481	0.870316176387881\\
12.3418227881859	0.921043824049406\\
12.4558303886926	0.972986714100944\\
12.5838389225948	1.02870020515332\\
12.7388492566171	1.09353682817941\\
12.9538635909061	1.18076380543025\\
13.5589039269285	1.42248185398797\\
15	1.99952044690471\\
};
\addplot [color=mycolor2, dashed, line width=1.5pt, forget plot]
  table[row sep=crcr]{%
0	0\\
4.99933328888593	1.99973331555422\\
5.00033335555704	0\\
9.99966664444296	1.99973331555422\\
10.0006667111141	0\\
15	1.99973331555422\\
};
\end{axis}

\begin{axis}[%
width=0.951\figurewidth,
height=0.419\figureheight,
at={(0\figurewidth,0\figureheight)},
scale only axis,
xmin=0,
xmax=15,
xlabel style={font=\color{white!15!black}},
xlabel={$t$},
ymin=-1.5,
ymax=2.5,
ylabel style={font=\color{white!15!black}},
axis background/.style={fill=white},
xmajorgrids,
ymajorgrids,
ylabel near ticks,
xlabel near ticks,
title style={font=\normalsize},
xlabel style={font=\footnotesize},
xticklabel style ={/pgf/number format/fixed, /pgf/number format/precision=3},
ylabel style={font=\footnotesize},
yticklabel style ={/pgf/number format/fixed, /pgf/number format/precision=3},
xlabel shift={-3pt},
ylabel shift={-5pt},
zlabel shift={-5pt},
legend style={font=\scriptsize},
ticklabel style={font=\scriptsize}
]
\addplot [color=mycolor1, line width=1.5pt, forget plot]
  table[row sep=crcr]{%
0	0\\
0.0700046669777983	0.00113475339099978\\
0.0980065337689187	0.00430080985029591\\
0.123008200546703	0.00961081580508427\\
0.148009867324488	0.0174977064517154\\
0.174011600773385	0.0283352841712468\\
0.203013534235616	0.0431477898222958\\
0.239015934395626	0.0644345566530884\\
0.383025535035669	0.152886320502803\\
0.423028201880125	0.172833117397975\\
0.464030935395693	0.190556119113609\\
0.51103406893793	0.208115452113153\\
0.57503833588906	0.229144801166477\\
0.728048536569105	0.2785598191817\\
0.813054203613575	0.309228999302839\\
0.925061670778051	0.352431261124217\\
1.12007467164478	0.430557132137922\\
1.31808787252483	0.512328053867583\\
2.0511367424495	0.818085974887039\\
2.41216081072072	0.964249204005824\\
3.16121074738316	1.26417719804457\\
5.01833455563704	2.00592135936102\\
5.0233348889926	2.00435796233293\\
5.02733515567705	2.0001718746231\\
5.03133542236149	1.99255569119939\\
5.03633575571705	1.97733266018859\\
5.04233615574372	1.94970633720763\\
5.0493366224415	1.90373128958142\\
5.05733715581039	1.83288865604836\\
5.06633775585039	1.73090273042275\\
5.07733848923262	1.57753126284308\\
5.09133942262818	1.3449724822012\\
5.11034068937929	0.983670649867811\\
5.17134475631709	-0.210877996773714\\
5.18834588972598	-0.484186553747085\\
5.20334688979265	-0.688317932802413\\
5.2163477565171	-0.834342523710113\\
5.22834855657044	-0.942611703682795\\
5.23834922328155	-1.01320712878688\\
5.24734982332155	-1.06163542863954\\
5.25535035669045	-1.09291969929628\\
5.26235082338823	-1.11146376973644\\
5.26835122341489	-1.12100371505944\\
5.27335155677045	-1.12461486311651\\
5.27835189012601	-1.12440864714662\\
5.28335222348157	-1.12051336168737\\
5.28835255683712	-1.11306514866839\\
5.29435295686379	-1.09963935777841\\
5.30135342356157	-1.07811621716696\\
5.31035402360157	-1.04183374209915\\
5.32035469031269	-0.991232067298379\\
5.33235549036602	-0.918041524775383\\
5.3473564904327	-0.81081092566224\\
5.36635775718381	-0.656510844571562\\
5.39735982398827	-0.381635878573334\\
5.43736249083272	-0.0307673970047446\\
5.46036402426829	0.150022758362461\\
5.4793652910194	0.281628536282668\\
5.4963664244283	0.383748564829565\\
5.51236749116608	0.46557129334661\\
5.52636842456164	0.525637494284185\\
5.53936929128609	0.571908137194121\\
5.55137009133942	0.606750346656945\\
5.56337089139276	0.634380159242458\\
5.57437162477498	0.653726796642875\\
5.5843722914861	0.666670480624051\\
5.5933728915261	0.674797845305184\\
5.60237349156611	0.679832384619727\\
5.61137409160611	0.682018284894463\\
5.62037469164611	0.681605317085934\\
5.63037535835722	0.678405344132516\\
5.64137609173945	0.671951843233657\\
5.65337689179279	0.661923307303304\\
5.66737782518835	0.647022174705118\\
5.68537902526835	0.624114136935592\\
5.71138075871725	0.586693596005194\\
5.76438429228615	0.509757699683709\\
5.78838589239283	0.47963634877086\\
5.80938729248617	0.457111868096076\\
5.82838855923728	0.440116397408346\\
5.84638975931729	0.427020150202759\\
5.86539102606841	0.416254007944653\\
5.88439229281952	0.408393409602981\\
5.90439362624175	0.402901344137051\\
5.92539502633509	0.399730947026891\\
5.94939662644176	0.398680492145354\\
5.97839855990399	0.399991559294934\\
6.0224014934329	0.404794927411411\\
6.1104073604907	0.414685334729199\\
6.17741182745516	0.419149401476488\\
6.26041736115741	0.424952689300209\\
6.30542036135742	0.430656603863969\\
6.34542302820188	0.438134992587381\\
6.38542569504634	0.448141522358169\\
6.4264284285619	0.461001258248368\\
6.47043136209081	0.477461433816551\\
6.51943462897527	0.498476734109024\\
6.57843856257084	0.526548924676081\\
6.65744382958864	0.567036799546004\\
6.79345289685979	0.639900408048042\\
7.04946996466431	0.777022515036714\\
7.17247816521101	0.839959547250857\\
7.28648576571772	0.895578620946639\\
7.40349356623775	0.949935766467499\\
7.53150210014001	1.00668250047873\\
7.6795119674645	1.06958084231032\\
7.86652443496233	1.14629936103357\\
8.16754450296686	1.26681864406395\\
9.11960797386492	1.64710887765715\\
10.017667844523	2.00586514803389\\
10.0226681778785	2.00493874900945\\
10.0276685112341	2.00015786950579\\
10.0316687779185	1.99254202054618\\
10.0366691112741	1.97731939867657\\
10.0426695113008	1.94969355375334\\
10.0496699779985	1.90371904638521\\
10.0576705113674	1.83287700774288\\
10.0666711114074	1.73089172334031\\
10.0776718447897	1.57752100036507\\
10.0916727781852	1.34496310744159\\
10.1106740449363	0.983662377598458\\
10.1716781118741	-0.210883443609863\\
10.188679245283	-0.484191385404321\\
10.2036802453497	-0.688322276843492\\
10.2166811120741	-0.834346484520259\\
10.2286819121275	-0.942615341428635\\
10.2386825788386	-1.01321051871966\\
10.2476831788786	-1.06163861144448\\
10.2556837122475	-1.0929227101391\\
10.2626841789453	-1.1114666391601\\
10.2686845789719	-1.12100646977491\\
10.2736849123275	-1.12461752670442\\
10.278685245683	-1.12441122356319\\
10.2836855790386	-1.12051585479572\\
10.2886859123942	-1.11306756223982\\
10.2946863124208	-1.09964168075619\\
10.3016867791186	-1.07811844094562\\
10.3106873791586	-1.04183584824221\\
10.3206880458697	-0.991234055243048\\
10.3326888459231	-0.918043387354308\\
10.3476898459897	-0.810812655199136\\
10.3666911127408	-0.656512440286148\\
10.3976931795453	-0.38163732946632\\
10.4376958463898	-0.0307687730874804\\
10.4606973798253	0.150021379065853\\
10.4796986465764	0.281627133608783\\
10.4966997799853	0.383747127721161\\
10.5127008467231	0.465569813839847\\
10.5267017801187	0.525635970855085\\
10.5397026468431	0.571906568081241\\
10.5517034468965	0.606748731773427\\
10.5637042469498	0.634378495634556\\
10.574704980332	0.65372508617118\\
10.5847056470431	0.66666872601566\\
10.5937062470831	0.674796049930816\\
10.6027068471231	0.679830547664221\\
10.6117074471631	0.682016405701479\\
10.6207080472031	0.681603395150463\\
10.6307087139143	0.67840337428804\\
10.6417094472965	0.671949820397556\\
10.6537102473498	0.661921226576352\\
10.6677111807454	0.647020026698595\\
10.6857123808254	0.624111903473008\\
10.7117141142743	0.586691242667701\\
10.7647176478432	0.509755122499019\\
10.7887192479499	0.479633682034478\\
10.8097206480432	0.457109130050911\\
10.8287219147943	0.440113600896355\\
10.8467231148743	0.427017303833098\\
10.8657243816254	0.416251114956619\\
10.8847256483766	0.408390476283905\\
10.9047269817988	0.402898375225488\\
10.9257283818921	0.399727948330376\\
10.9497299819988	0.398677468859825\\
10.978731915461	0.399988519506527\\
11.0227348489899	0.404791889138119\\
11.1107407160477	0.414682386462641\\
11.1777451830122	0.419146587037364\\
11.2607507167144	0.424950099012337\\
11.3057537169145	0.430654154486231\\
11.3457563837589	0.438132676217313\\
11.3857590506034	0.448139343908441\\
11.4267617841189	0.460999224065164\\
11.4707647176478	0.477459555410235\\
11.5197679845323	0.498475027690198\\
11.5787719181279	0.526547419696902\\
11.6577771851457	0.567035548725501\\
11.7937862524168	0.639899537854371\\
12.0498033202213	0.777022137877688\\
12.1728115207681	0.839959312303295\\
12.2868191212748	0.895578475123386\\
12.4038269217948	0.949935679444526\\
12.531835455697	1.00668245072865\\
12.6798453230215	1.06958081318056\\
12.8668577905194	1.14629934013861\\
13.1678778585239	1.26681862453206\\
14.119941329422	1.64710887284188\\
15	1.9995154544911\\
};
\addplot [color=mycolor2, dashed, line width=1.5pt, forget plot]
  table[row sep=crcr]{%
0	0\\
4.99933328888593	1.99973331555422\\
5.00033335555704	0\\
9.99966664444296	1.99973331555422\\
10.0006667111141	0\\
15	1.99973331555422\\
};
\end{axis}
\end{tikzpicture}%

%% file: distBehavior.tex
%
%
\definecolor{mycolor1}{rgb}{0.00000,0.44700,0.74100}%
\begin{tikzpicture}

\begin{axis}[%
width=0.951\figurewidth,
height=0.419\figureheight,
at={(0\figurewidth,0.581\figureheight)},
scale only axis,
xmin=0,
xmax=5,
ymin=-0.055,
ymax=0.075,
ylabel style={font=\color{white!15!black}},
xticklabels={\empty},
axis background/.style={fill=white},
xmajorgrids,
ymajorgrids,
ylabel near ticks,
xlabel near ticks,
title style={font=\normalsize},
xlabel style={font=\footnotesize},
xticklabel style ={/pgf/number format/fixed, /pgf/number format/precision=3},
ylabel style={font=\footnotesize},
yticklabel style ={/pgf/number format/fixed, /pgf/number format/precision=3},
ylabel shift={-5pt},
zlabel shift={-5pt},
legend style={font=\scriptsize},
ticklabel style={font=\scriptsize}
]
\addplot [color=mycolor1, line width=1.5pt, forget plot]
  table[row sep=crcr]{%
0	0\\
0.00300020001333401	4.27582114843617e-05\\
0.00600040002666802	0.000172409860708989\\
0.00900060004000292	0.000390979320715523\\
0.0120008000533369	0.000700013907077413\\
0.0150010000666709	0.00109989431932256\\
0.0180012000800049	0.00158948890230981\\
0.0210014000933398	0.00216620587331828\\
0.0240016001066738	0.00282623304588725\\
0.0270018001200079	0.00356481971724421\\
0.0300020001333419	0.00437653757446377\\
0.0330022001466768	0.00525550079318116\\
0.0360024001600108	0.00619554371455688\\
0.0390026001733448	0.00719036092843428\\
0.0430028668577904	0.00859104298008884\\
0.0470031335422361	0.0100629995650365\\
0.0520034668977933	0.0119812523136407\\
0.0580038669244614	0.0143627956832759\\
0.0700046669777983	0.0192263579182246\\
0.0780052003466896	0.0224263110113982\\
0.0840056003733585	0.0247521027152899\\
0.0890059337289149	0.0266210838937422\\
0.0940062670844721	0.0284137441219938\\
0.0980065337689178	0.0297856174597122\\
0.102006800453363	0.0310969340800149\\
0.106007067137809	0.0323435406524819\\
0.110007333822255	0.0335218718772605\\
0.1140076005067	0.0346289154717461\\
0.118007867191146	0.0356621777975388\\
0.122008133875592	0.0366196503137299\\
0.126008400560037	0.037499776991333\\
0.130008667244483	0.0383014227847287\\
0.134008933928929	0.0390238432262349\\
0.138009200613374	0.0396666551871077\\
0.14200946729782	0.0402298088306745\\
0.146009733982265	0.040713560769686\\
0.150010000666711	0.0411184484293523\\
0.154010267351157	0.041445265609263\\
0.158010534035602	0.0416950392308584\\
0.162010800720048	0.0418690072520338\\
0.166011067404494	0.0419685977264157\\
0.170011334088939	0.0419954089816938\\
0.174011600773385	0.0419511908889225\\
0.178011867457831	0.0418378271927731\\
0.182012134142276	0.0416573188712794\\
0.186012400826722	0.0414117684924831\\
0.190012667511168	0.0411033655346484\\
0.194012934195613	0.0407343726361589\\
0.198013200880059	0.0403071127409156\\
0.202013467564504	0.0398239571049448\\
0.20601373424895	0.0392873141299503\\
0.211014067604507	0.0385450054393646\\
0.216014400960064	0.0377277204344448\\
0.221014734315621	0.0368402657432085\\
0.226015067671178	0.0358874389241528\\
0.231015401026736	0.0348740066694262\\
0.236015734382292	0.0338046853181613\\
0.242016134408961	0.0324542657253444\\
0.248016534435629	0.0310379281224709\\
0.255017001133409	0.029312438305773\\
0.262017467831189	0.0275194271763377\\
0.27001800120008	0.0254020568594084\\
0.279018601240082	0.0229521312599239\\
0.290019334622309	0.0198901233550037\\
0.310020668044537	0.0142408784256718\\
0.326021734782318	0.00974865339769249\\
0.337022468164545	0.00671946399618939\\
0.347023134875658	0.0040284543417819\\
0.356023734915661	0.00166937666239253\\
0.365024334955664	-0.000621329253684522\\
0.373024868324555	-0.00259422474249948\\
0.381025401693447	-0.00450303240215177\\
0.389025935062338	-0.00634411134735213\\
0.397026468431228	-0.00811445316050641\\
0.404026935129009	-0.00960355817140357\\
0.411027401826789	-0.0110353618229961\\
0.418027868524568	-0.0124088778828604\\
0.425028335222348	-0.0137233927119587\\
0.432028801920128	-0.0149784439671032\\
0.439029268617908	-0.0161738000919973\\
0.446029735315688	-0.0173094406516636\\
0.453030202013467	-0.0183855375504649\\
0.460030668711248	-0.0194024371611388\\
0.467031135409028	-0.0203606433809629\\
0.474031602106807	-0.0212608016214038\\
0.481032068804587	-0.0221036837291235\\
0.488032535502366	-0.0228901738289764\\
0.496033068871258	-0.023721233468879\\
0.504033602240149	-0.0244815300842971\\
0.51203413560904	-0.0251727941195439\\
0.520034668977932	-0.0257968611538599\\
0.528035202346823	-0.0263556531974265\\
0.536035735715714	-0.0268511616666851\\
0.544036269084605	-0.0272854319495108\\
0.552036802453497	-0.0276605494685898\\
0.560037335822388	-0.0279786271504996\\
0.569037935862391	-0.0282709355394362\\
0.578038535902394	-0.0284967881437419\\
0.587039135942396	-0.0286592195669018\\
0.596039735982399	-0.0287612449613066\\
0.606040402693513	-0.0288073877925488\\
0.616041069404627	-0.0287866704940516\\
0.626041736115742	-0.0287030481533996\\
0.636042402826855	-0.0285603867740205\\
0.64704313620908	-0.0283397625397068\\
0.658043869591306	-0.0280571235783338\\
0.670044669644643	-0.0276835833697247\\
0.68204546969798	-0.0272477064985015\\
0.695046336422428	-0.0267116805907808\\
0.709047269817987	-0.0260678016777121\\
0.724048269884659	-0.0253102055921142\\
0.740049336622442	-0.0244351578363204\\
0.757050470031335	-0.0234412864996631\\
0.776051736782453	-0.0222664697962633\\
0.798053203546903	-0.0208408252902457\\
0.825055003666911	-0.0190248946134757\\
0.869057937195813	-0.0159915913624493\\
0.917061137409161	-0.0127006166128716\\
0.947063137542503	-0.0107064957110161\\
0.9730648709914	-0.00903923200414081\\
0.997066471098074	-0.00756137077612884\\
1.02006800453364	-0.00620736451774739\\
1.04206947129809	-0.00497421148758992\\
1.06407093806254	-0.00380531105530313\\
1.08507233815588	-0.00275194566842352\\
1.10607373824922	-0.00176119172013589\\
1.12707513834256	-0.000834057204374794\\
1.1480765384359	2.90065615518031e-05\\
1.16907793852924	0.000828047353715\\
1.19007933862258	0.00156355766718264\\
1.21108073871591	0.00223642346189035\\
1.23208213880925	0.00284787595432956\\
1.25408360557371	0.00342424750256676\\
1.27608507233816	0.00393696327456983\\
1.29808653910261	0.00438834686749878\\
1.32108807253817	0.00479742799172111\\
1.34408960597373	0.00514535112205117\\
1.36809120608041	0.00544672239056521\\
1.39309287285819	0.00569774522759925\\
1.41909460630709	0.00589541586409936\\
1.4460964064271	0.00603760361973293\\
1.47409827321821	0.0061231131006183\\
1.50410027335156	0.00615170454226188\\
1.53610240682712	0.00611854553704738\\
1.57010467364491	0.00602086318399664\\
1.60810720714714	0.00584753010453909\\
1.65011000733382	0.00559222903960688\\
1.69911327421828	0.00523047292394452\\
1.7621174744983	0.00469861865171328\\
1.88912594172945	0.00354429460483363\\
1.98213214214281	0.00273518851428545\\
2.05513700913394	0.00216280415600067\\
2.12214147609841	0.00169932621110114\\
2.18814587639176	0.00130526961280797\\
2.25415027668511	0.000973457135476252\\
2.32215481032069	0.000693753795522944\\
2.39415961064071	0.000460540140138299\\
2.4721648109874	0.000271923093956516\\
2.55817054470298	0.000128266659339182\\
2.6561770784719	2.86537926692532e-05\\
2.77618507900527	-2.7927666851113e-05\\
2.94119607973865	-3.81675594756103e-05\\
4.08627241816121	1.55668201298198e-05\\
4.99933328888593	2.78029487432718e-07\\
};
\end{axis}

\begin{axis}[%
width=0.951\figurewidth,
height=0.419\figureheight,
at={(0\figurewidth,0\figureheight)},
scale only axis,
xmin=0,
xmax=5,
xlabel style={font=\color{white!15!black}},
xlabel={$t$},
ymin=-0.055,
ymax=0.075,
ylabel style={font=\color{white!15!black}},
axis background/.style={fill=white},
xmajorgrids,
ymajorgrids,
ylabel near ticks,
xlabel near ticks,
title style={font=\normalsize},
xlabel style={font=\footnotesize},
xticklabel style ={/pgf/number format/fixed, /pgf/number format/precision=3},
ylabel style={font=\footnotesize},
yticklabel style ={/pgf/number format/fixed, /pgf/number format/precision=3},
xlabel shift={-3pt},
ylabel shift={-5pt},
zlabel shift={-5pt},
legend style={font=\scriptsize},
ticklabel style={font=\scriptsize}
]
\addplot [color=mycolor1, line width=1.5pt, forget plot]
  table[row sep=crcr]{%
0	0\\
0.00300020001333401	4.26553591390544e-05\\
0.00600040002666802	0.00017157988857619\\
0.00900060004000292	0.000388219346151963\\
0.0120008000533369	0.000693970025366752\\
0.0150010000666709	0.00108997108540532\\
0.0180012000800049	0.0015768310101576\\
0.0210014000933398	0.00215443743174504\\
0.0240016001066738	0.00282188833505082\\
0.0270018001200079	0.0035775126504447\\
0.0300020001333419	0.00441894034776702\\
0.0330022001466768	0.00534319361271152\\
0.0360024001600108	0.00634678239976338\\
0.0390026001733448	0.00742579566300439\\
0.0430028668577904	0.00897442351279132\\
0.0470031335422361	0.0106387810347384\\
0.0510034002266817	0.0124075173760492\\
0.0550036669111273	0.0142689204248274\\
0.0600040002666846	0.0167078544156922\\
0.0650043336222419	0.0192496804257027\\
0.0710047336489099	0.022403224186089\\
0.0790052670178012	0.0267197042075997\\
0.0970064670978061	0.0364806579868961\\
0.103006867124475	0.0396322252592096\\
0.109007267151143	0.0426838732589641\\
0.1140076005067	0.0451319614982086\\
0.119007933862258	0.0474791835997159\\
0.124008267217815	0.0497131453834756\\
0.128008533902261	0.0514111893983369\\
0.132008800586706	0.0530243369236141\\
0.136009067271152	0.0545478499577854\\
0.140009333955597	0.0559774528939041\\
0.144009600640043	0.0573093241166944\\
0.148009867324489	0.0585400866723855\\
0.152010134008934	0.0596667981174059\\
0.15601040069338	0.0606869396404184\\
0.160010667377825	0.0615984045425106\\
0.16401093406227	0.0623994861524277\\
0.168011200746716	0.0630888652470372\\
0.172011467431162	0.0636655970416182\\
0.176011734115607	0.064129097809789\\
0.180012000800053	0.0644791311887651\\
0.184012267484499	0.0647157942220513\\
0.188012534168944	0.0648395031885203\\
0.19201280085339	0.0648509792639826\\
0.196013067537836	0.0647512340587975\\
0.200013334222281	0.0645415550727328\\
0.204013600906727	0.0642234911060635\\
0.208013867591172	0.063798837663902\\
0.212014134275618	0.0632696223887628\\
0.216014400960064	0.0626380905545387\\
0.220014667644509	0.061906690653287\\
0.224014934328955	0.0610780601044754\\
0.228015201013401	0.0601550111146922\\
0.232015467697846	0.0591405167141765\\
0.236015734382292	0.0580376969949326\\
0.240016001066738	0.0568498055736422\\
0.244016267751183	0.0555802163010322\\
0.248016534435629	0.0542324102378933\\
0.253016867791186	0.052443115382558\\
0.258017201146743	0.0505443599467323\\
0.2630175345023	0.0485434678139027\\
0.268017867857857	0.046447890803953\\
0.273018201213414	0.0442651744474318\\
0.279018601240082	0.0415416031254328\\
0.285019001266751	0.0387167001368551\\
0.291019401293419	0.0358036470035197\\
0.2980198679912	0.0323111412368622\\
0.306020401360091	0.0282209441724266\\
0.316021068071205	0.0230038898246097\\
0.336022401493433	0.0124337763272067\\
0.34802320154677	0.00614408627221774\\
0.357023801586773	0.00151364853678793\\
0.365024334955664	-0.00251197780310086\\
0.372024801653444	-0.00594850814028991\\
0.379025268351223	-0.00929228127887338\\
0.386025735049003	-0.0125324018418347\\
0.392026135075672	-0.0152196883989939\\
0.39802653510234	-0.0178180669354857\\
0.404026935129009	-0.0203225884228369\\
0.410027335155677	-0.0227288835640085\\
0.416027735182346	-0.0250331540631006\\
0.422028135209014	-0.0272321615457267\\
0.428028535235683	-0.0293232143250828\\
0.433028868591239	-0.0309817188433819\\
0.438029201946796	-0.0325627692234765\\
0.443029535302354	-0.0340656298739228\\
0.448029868657911	-0.0354897980880926\\
0.453030202013467	-0.0368349957362781\\
0.458030535369025	-0.0381011605822605\\
0.463030868724582	-0.039288437273127\\
0.468031202080138	-0.0403971680497746\\
0.473031535435696	-0.0414278832241433\\
0.478031868791253	-0.0423812914677439\\
0.48303220214681	-0.0432582699544195\\
0.488032535502366	-0.0440598543986805\\
0.493032868857924	-0.044787229029212\\
0.499033268884593	-0.0455639931317755\\
0.505033668911261	-0.0462383343541539\\
0.51103406893793	-0.046813021922425\\
0.517034468964598	-0.047291041870686\\
0.523034868991266	-0.0476755727727118\\
0.529035269017935	-0.0479699619236547\\
0.535035669044603	-0.0481777020626595\\
0.541036069071271	-0.0483024087191524\\
0.54703646909794	-0.048347798257506\\
0.55403693579572	-0.0483055724475152\\
0.5610374024935	-0.0481666835323136\\
0.568037869191279	-0.047937326564675\\
0.575038335889059	-0.0476237249370461\\
0.58303886925795	-0.0471701366288002\\
0.591039402626842	-0.0466238332114672\\
0.599039935995733	-0.045993881703752\\
0.608040536035736	-0.0451962412048328\\
0.61804120274685	-0.0442138743059131\\
0.628041869457964	-0.0431459318435472\\
0.640042669511301	-0.0417729326687777\\
0.65404360290686	-0.0400759762261167\\
0.672044802986866	-0.0377977849640061\\
0.730048669911327	-0.0303841393945392\\
0.747049803320222	-0.0283320506760578\\
0.763050870058004	-0.0264864719510456\\
0.778051870124675	-0.024839830872212\\
0.793052870191346	-0.0232776522067137\\
0.808053870258017	-0.0218003291213513\\
0.823054870324689	-0.0204057577409831\\
0.839055937062471	-0.0190048105638247\\
0.856057070471365	-0.0176067945198062\\
0.87405827055137	-0.0162176307133572\\
0.893059537302487	-0.0148394536196523\\
0.914060937395827	-0.0134040918490381\\
0.9380625375025	-0.0118533783791408\\
0.966064404293619	-0.0101342220519074\\
1.00006667111141	-0.00813718507543459\\
1.04006933795586	-0.00587628021221143\\
1.08007200480032	-0.00369948348072224\\
1.11507433828922	-0.00187768325548632\\
1.14607640509367	-0.000348905306108094\\
1.17407827188479	0.000946336105740464\\
1.20008000533369	0.00206358102218118\\
1.22508167211147	0.00305098619960908\\
1.24908327221815	0.00391252383026153\\
1.27308487232482	0.00468548407492619\\
1.29608640576038	0.00534133886296306\\
1.31908793919595	0.00591407172208402\\
1.34308953930262	0.0064246966575654\\
1.36709113940929	0.00684959688863795\\
1.39209280618708	0.00720579332960281\\
1.41809453963598	0.00748910595883867\\
1.44509633975598	0.0076972999519711\\
1.47409827321821	0.00783335767784799\\
1.50510034002267	0.00789075037210374\\
1.53910260684046	0.00786410335350318\\
1.57610507367158	0.0077463351077105\\
1.61710780718715	0.00752851121305032\\
1.66411094072938	0.00719093705057894\\
1.72011467431162	0.00669854850498641\\
1.79011934128942	0.00599172273129778\\
1.90812720848057	0.00469774423879166\\
2.02813520901393	0.00341196324941162\\
2.10814054270285	0.0026412531566038\\
2.18114540969398	0.00202354750648137\\
2.25215014334289	0.00150915511308281\\
2.32415494366291	0.0010750272098532\\
2.39815987732516	0.000715944092331888\\
2.47616507767184	0.00042403649360967\\
2.56117074471632	0.000193641657658006\\
2.65517701180079	2.65737433799984e-05\\
2.76418427895193	-7.88114135099605e-05\\
2.90119341289419	-0.000120317203229803\\
3.10820721381425	-8.64594072469416e-05\\
3.68324554970331	3.64367494150031e-05\\
4.22328155210347	2.40674317328882e-05\\
4.99933328888593	5.50212677197237e-07\\
};
\end{axis}
\end{tikzpicture}%